\documentclass[aps,pre,reprint,superscriptaddress,amsmath,amssymb,nofootinbib,floatfix]{revtex4-2}
\usepackage{graphicx,booktabs,bm}
\usepackage{microtype}
\usepackage[colorlinks=true,linkcolor=blue,citecolor=blue,urlcolor=blue]{hyperref}
\makeatletter
\AtBeginDocument{\let\auto@bib@innerbib\@empty
  \let\write@bibliographystyle\relax}
\makeatother
\newcommand{\Dr}{D[p(t)\Vert\pi_b]}

\hypersetup{pdftitle={Mpemba effect as an entropy-production race},pdfauthor={Mesfin Asfaw Taye}}

\begin{document}

\title{Mpemba effect as an entropy-production race:\texorpdfstring{\\}{ }
Finite-amplitude bounds on crossing times in Markovian relaxation}

\author{Mesfin Asfaw Taye}
\email{tayem@wlac.edu}
\affiliation{West Los Angeles College, Science Division,
9000 Overland Avenue, Culver City, California 90230, USA}

\date{October 24, 2026}

\begin{abstract}
Two canonical preparations evolving under a common final Markov generator
can exhibit the Mpemba effect because they excite different relaxation
modes. We bound the first crossing of their relative entropies by treating
relaxation as an entropy-production race. The exact balance
$-\dot D=\dot e_p$ implies that a meeting occurs when the accumulated
\emph{difference} of normalized entropy-production rates,
$\int_0^t(r_H-r_C)\,\mathrm ds$ with $r_\alpha=\dot e_p^{\alpha}/D_\alpha$,
equals the initial logarithmic deficit $\ln[D_H(0)/D_C(0)]$.
Global entropy-decay inequalities yield a lower crossing-time bound when
the entropy-decay constant is bounded from below and the hot rate from
above; we supply both certificates. A nonzero cubic moment of a slowest
eigenfunction is sufficient for the modified logarithmic Sobolev constant
to lie strictly below the spectral gap. Combining established relative-entropy
comparison inequalities with Markov contraction and the actual mode
amplitudes gives finite-amplitude crossing bounds without a small-deviation
expansion. For an exactly strong, reversible three-state preparation these
give $0.1659\le t_M\le0.1674$ in model time units, independently certified
by outward-rounded interval arithmetic. This interval is narrower than
$1\%$ and contains $t_M=0.1662209$. The conservative global-rate comparison
used here gives only $t_M\ge0.042$. The hotter preparation starts $6.12$
times farther from equilibrium, yet it reaches $D=10^{-5}$ at $t=0.233$,
compared with $t=0.844$ for the colder one. Mean energy crosses later than
relative entropy.
For a fixed driven generator, nonadiabatic entropy production governs the
same balance. These results concern finite-state relaxation in a specified
distance and do not predict water-freezing times.
\end{abstract}

\maketitle

\section{Introduction}
\label{sec:intro}

The farther a system starts from equilibrium, the longer it should take to
get there. The Mpemba effect---the observation that hot water can freeze
before cold water \cite{mpemba1969,jeng2006}---challenges this intuition.
In water, however, the effect is entangled with evaporation, convection,
supercooling, and nucleation, and its reproducibility has been debated
\cite{jeng2006,burridge2016}. Over the past decade the question has been
sharpened into a clean problem of nonequilibrium statistical mechanics
\cite{teza2026}: two copies of the same system, equilibrated at
\begin{equation}
T_H>T_C>T_b,
\label{eq:temps}
\end{equation}
are quenched into one bath at $T_b$ and thereafter obey identical dynamics.
Can the hotter copy reach equilibrium first?

For Markovian dynamics the answer is yes, and the mechanism is spectral
\cite{luraz2017,klich2019}. Both copies share the post-quench generator
and therefore its relaxation rates; they differ only in how their initial
states project onto its modes. A hot preparation with a weaker overlap
with the slowest mode can overtake the cold one, and when that overlap
vanishes the hotter system relaxes exponentially faster---the strong
Mpemba effect \cite{klich2019}. Anomalous relaxation of this kind has been
predicted in granular fluids \cite{lasanta2017,biswas2020}, spin glasses
\cite{baityjesi2019}, Langevin dynamics \cite{walker2021}, and open and
many-body quantum systems \cite{carollo2021,ares2023,moroder2024}, and it
has been observed in colloids \cite{kumar2020,kumar2022} and trapped ions
\cite{joshi2024,shapira2024}. Inverse effects, precooling protocols, and
protocols that include the preparation time
\cite{luraz2017,gal2020,kumar2022,nava2025}, as well as heating--cooling
asymmetries between thermodynamically equidistant quenches
\cite{lapolla2020,vuhasegawa2021,meibohm2021,ibanez2024}, belong to a
broader family of relaxation speedups \cite{teza2026}.

What remains open is the timing. An asymptotic mode-ordering criterion
tells whether two trajectories eventually exchange order, but not when
they first meet. To locate that meeting we use a thermodynamic distance.
The relative entropy
$D[p\Vert\pi_b]$ to the final state equals the excess nonequilibrium free
energy in units of $k_{\rm B}T_b$, and its rate of decrease is exactly the
entropy-production rate $\dot e_p$
\cite{schnakenberg1976,seifert2005,seifert2012,esposito2010}. Entropy
production is the organizing quantity of stochastic thermodynamics
\cite{Sekimoto2010,seifert2012,vankampen2007}, from Brownian motors and
heat engines
\cite{Reimann2002,Hanggi2009,Landauer1975,Buttiker1987,Derenyi1999,Hondou2000,CurzonAhlborn1975,Berger2009,Polettini2013,Celani2012,BoCelani2013,Asfaw2002AdjustableBrownianHeat,AsfawBekele2004,Asfaw2005EnergeticsSimpleMicroscopic,Asfaw2007ExploringOperationTiny,Asfaw2008ModelingEfficientBrownian,Asfaw2013EffectThermalInhomogeneity,Asfaw2014ThermodynamicFeatureBrownian,Taye2015EffectTemperatureDependence,Duki2015EffectTemperatureViscous,Taye2017,Taye2021BrownianMotorsArranged,Taye2025ThermodynamicFeaturesHeat,TayeExp2025,Taye2025CompetingActivePassive,Taye2026ExactThermodynamicAnalysis,Taye2026BrownianMotorsBrownian}
to exact time-dependent entropy-production and entropy-extraction rates of
Brownian particles in overdamped and underdamped media
\cite{Taye2015ExactAnalyticalThermodynamic,Taye2015ExactAnalyticalExpressions,Taye2016,Taye2020,Taye2021,Taye2022ExactTimeDependent2,Taye2023TimeDependentSolutions,Taye2024ExactTimeDependent,Taye2024ExactTimeDependent2,Taye2025ThermodynamicIrreversibilityUnderdamped,Taye2025EntropyProductionThermodynamic,taye2025EntropyProductionThermodynamic,Taye2025ThermodynamicRelationsTime,Taye2025UnifiedNonequilibriumFramework,Taye2026UniversalThermodynamicInequality,Taye2026NonequilibriumThermodynamicsStochastic,Taye2026EntropyProductionMacroscopic}.
Entropy production has also been tied to relaxation durations through
functional inequalities \cite{bobkov2006,diaconis1996}: Bao et al. relate
it to logarithmic Sobolev bounds on relaxation duration \cite{bao2025},
and Auconi discusses such inequalities for instantaneous relaxation in
diffusion processes \cite{auconi2026}. A global constant holds for every initial
state; by itself it does not retain the particular small slow-mode
amplitude that can govern a crossing. Thermomajorization \cite{vu2025} and
resource-theoretic approaches \cite{summer2026} address a complementary
question, namely whether an ordering holds simultaneously for all
monotones.

Here we express the Mpemba meeting as an entropy-production race with an
exact balance and obtain a tight crossing interval for a specified pair.
Normalizing $\dot e_p$ by
$D$ gives the fraction of the remaining excess free energy dissipated per
unit time. The relative entropies meet when the hot copy's accumulated
advantage in this normalized rate equals its initial logarithmic deficit
$\ln[D_H(0)/D_C(0)]$. Three results follow. (i)~Global entropy-decay
inequalities convert this identity into a crossing-time bound only if the
decay constant is bounded from below and the hot rate from above; we
provide an analytic reset certificate for the former and determine the
latter exactly for the strong preparation. (ii)~A nonzero third moment of
the slowest eigenfunction forces the entropy-decay (modified logarithmic
Sobolev) constant strictly below the spectral gap, so the near-equilibrium
spectral picture overestimates the uniform decay rate. (iii)~Retaining the
actual mode amplitudes, established finite-amplitude entropy comparisons
combined with Markov contraction, without quadratic truncation, confine
the crossing of a three-state
strong-Mpemba pair to $0.1659\le t_M\le0.1674$, a window narrower than
$1\%$ around the computed $t_M=0.16622$, compared with $t_M\ge0.042$ from
the conservative certified global-rate comparison used here. An independent
interval calculation certifies the numerical sign conditions and displayed
enclosure for the exact model. The
entropy-dissipation identity, entropy-decay inequality, and static
relative-entropy comparison are established results. Our contribution is
their preparation-dependent dynamical application: explicit lower and
upper first-crossing bounds, an iterative refinement, and a worked
finite-amplitude enclosure. Section~\ref{sec:envelopes} identifies the
static comparison precisely.

The analysis also clarifies what the effect is not. In our example the
mean energy crosses later than the relative entropy, so ``which system is
closer to equilibrium'' depends on the observable. For relaxation toward a
nonequilibrium steady state the same race is run with the nonadiabatic
entropy production \cite{esposito2010,hatano2001}, and housekeeping
dissipation drops out of the balance at fixed driving. None of these
statements concerns freezing water. They isolate the shared-generator,
preparation-dependent mechanism within the finite-state Markov framework.

\section{Two thermal preparations and one final bath}
\label{sec:setup}

\subsection{Finite-state Markov dynamics}

Consider a finite set of mesoscopic states $i=1,\dots,N$ with fixed energies
$E_i$ and standard continuous-time Markov dynamics
\cite{vankampen2007,norris1997}.  Let $k_{ij}$ denote the rate for a jump $i\to j$ after the quench.  The
probability $p_i(t)$ evolves according to
\begin{equation}
\dot p_i(t)=\sum_{j\ne i}\left[k_{ji}p_j(t)-k_{ij}p_i(t)\right].
\label{eq:master}
\end{equation}
The bath temperature after the quench is $T_b$; throughout most of the paper we
set $k_{\rm B}=1$.  The rates satisfy detailed balance with respect to
\begin{equation}
\pi_{b,i}=\frac{e^{-\beta_bE_i}}{Z_b},\qquad
\beta_b=\frac{1}{T_b},\qquad
Z_b=\sum_ie^{-\beta_bE_i},
\label{eq:gibbs}
\end{equation}
namely
\begin{equation}
\pi_{b,i}\,k_{ij}=\pi_{b,j}\,k_{ji}.
\label{eq:db}
\end{equation}
The process is assumed irreducible, so $\pi_b$ is the unique stationary
state. Every allowed edge has its reverse. Entropies are dimensionless:
the physical entropy-production rate is $k_{\rm B}\dot e_p$, and $\dot e_p$ has
units of inverse time. Rates and temperatures are fixed after the quench.

\begin{figure*}[!t]
\centering
\includegraphics[width=0.78\textwidth]{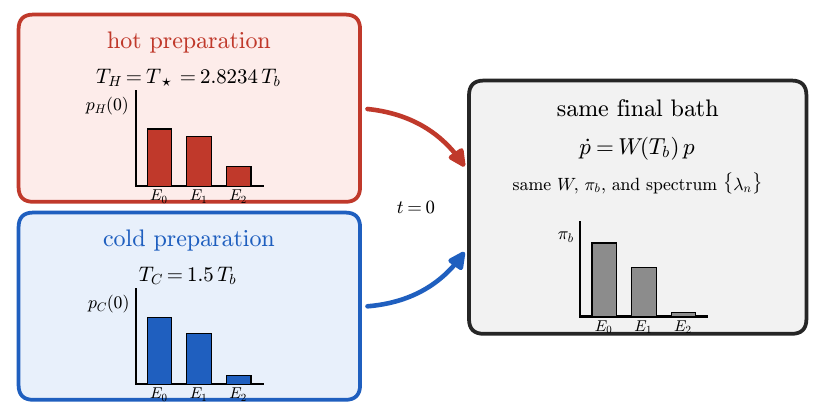}
\caption{Common-bath formulation of the Mpemba problem. Two copies of
the same three-state system are prepared in canonical states at
$T_H=T_\star$ and $T_C=1.5\,T_b$ (bar heights are the actual occupations of
the levels $E_0,E_1,E_2$). At $t=0$ both are quenched into one bath and
evolve with the same generator $W(T_b)$ toward the same equilibrium
distribution $\pi_b$. The spectrum is therefore common to the two copies;
preparation-dependent relaxation arises only through the different
projections of the initial states onto the common relaxation modes.}
\label{fig:schematic}
\end{figure*}

Two copies of the system are prepared before the quench in canonical
equilibrium at temperatures $T_H$ and $T_C$:
\begin{equation}
p^H_i(0)=\frac{e^{-E_i/T_H}}{Z(T_H)},\qquad
p^C_i(0)=\frac{e^{-E_i/T_C}}{Z(T_C)}.
\label{eq:preps}
\end{equation}
At $t=0$ both copies are connected to the same bath $T_b$ and subsequently
evolve with the same rates $k_{ij}$, as illustrated in
Fig.~\ref{fig:schematic}.  The quench protocol is therefore
\begin{equation}
\dot p^{\alpha}=W(T_b)\,p^{\alpha},\qquad \alpha\in\{H,C\},
\label{eq:protocol}
\end{equation}
with identical $W(T_b)$ for the two preparations.

\subsection{Nonequilibrium free energy and relative entropy}

The relative entropy with respect to the final equilibrium state is
\begin{equation}
D(t)\equiv\Dr=\sum_ip_i(t)\ln\frac{p_i(t)}{\pi_{b,i}}.
\label{eq:Ddef}
\end{equation}
It has a direct thermodynamic meaning.  Define the nonequilibrium free energy
\begin{equation}
\mathcal F[p]=\sum_ip_iE_i+T_b\sum_ip_i\ln p_i.
\label{eq:Fdef}
\end{equation}
Then
\begin{equation}
\mathcal F[p]-F_b=T_b\,D[p\Vert\pi_b],
\label{eq:FD}
\end{equation}
where $F_b=-T_b\ln Z_b$.  Thus $D$ measures the excess free energy available
for dissipation during isothermal relaxation.

For a canonical initial state at inverse temperature $\beta_0$, the initial
thermodynamic distance is known from equilibrium quantities alone:
\begin{equation}
D_0(\beta_0)=(\beta_b-\beta_0)\,U(\beta_0)+\ln\frac{Z(\beta_b)}{Z(\beta_0)},
\label{eq:D0}
\end{equation}
where $U(\beta_0)=\sum_iE_i\pi_i(\beta_0)$. Since
$\mathrm dU/\mathrm d\beta_0=-\operatorname{Var}_{T_0}(E)$ and
$\mathrm d\ln Z/\mathrm d\beta_0=-U$, differentiating gives
\begin{equation}
\frac{\mathrm dD_0}{\mathrm dT_0}
=\frac{\beta_b-\beta_0}{T_0^2}\operatorname{Var}_{T_0}(E)>0
\quad (T_0>T_b).
\label{eq:D0monotone}
\end{equation}
The inequality is strict for nonconstant energies. Thus the initial
ordering $D_H(0)>D_C(0)$ follows from the canonical cooling protocol itself.
Away from equilibrium, a distribution generally has no single temperature;
relative-entropy relaxation must be distinguished from the decay of mean
energy or a measured temperature.

\subsection{Entropy production}

For a pair of states define the probability current
\begin{equation}
J_{ij}(t)=k_{ij}p_i(t)-k_{ji}p_j(t)=-J_{ji}(t).
\label{eq:current}
\end{equation}
The total entropy-production rate is the Schnakenberg form
\cite{schnakenberg1976,seifert2005,seifert2012}
\begin{equation}
\dot e_p(t)=\frac12\sum_{i,j}J_{ij}(t)\,
\ln\frac{k_{ij}p_i(t)}{k_{ji}p_j(t)}\;\ge0.
\label{eq:sigma}
\end{equation}
Splitting the logarithm as $\ln(p_i/p_j)+\ln(k_{ij}/k_{ji})$ gives the
entropy balance
\begin{equation}
\dot S(t)=\dot e_p(t)-\dot h_d(t),\qquad
\dot h_d(t)=\frac12\sum_{i,j}J_{ij}(t)\ln\frac{k_{ij}}{k_{ji}},
\label{eq:balanceS}
\end{equation}
where $S=-\sum_ip_i\ln p_i$ is the system entropy, $\dot e_p\ge0$ is the
entropy-production rate, and $\dot h_d$ is the entropy-extraction
(heat-dissipation) rate into the bath. Under detailed balance,
$\ln(k_{ij}/k_{ji})=(E_i-E_j)/T_b$, so $\dot h_d=-\dot U/T_b$ is the heat
flow into the bath divided by its temperature.
Because $D=-S+U/T_b+\ln Z_b$, Eq.~\eqref{eq:balanceS} yields the exact identity
\begin{equation}
\dot e_p(t)=-\frac{\mathrm d}{\mathrm dt}\,\Dr.
\label{eq:identity}
\end{equation}
A direct derivation from the master equation is given in Appendix~\ref{app:identity}. Equation~\eqref{eq:identity} is
well known in stochastic thermodynamics---it expresses the statement
$\dot{\mathcal F}=-T_b\dot e_p$ for isothermal relaxation
\cite{seifert2012}---and it is the central thermodynamic simplification of the
common-bath problem: the decrease of excess nonequilibrium free energy is
exactly the irreversible entropy production.  It is special to relaxation
toward a detailed-balance equilibrium state; for a generic nonequilibrium
steady state, relative-entropy decay and total entropy production do not
coincide in this simple form \cite{maes2008}, a situation we return to in
Sec.~\ref{sec:ness}.

\section{Four relaxation times}
\label{sec:times}

The phrase ``relaxation time'' can refer to several distinct quantities.
Keeping them separate is essential when comparing hot and cold preparations.

\subsection{Spectral relaxation time}

Let the eigenvalues of the final generator be
\begin{equation}
0=-\lambda_1>-\lambda_2\ge-\lambda_3\ge\cdots,\qquad
0<\lambda_2\le\lambda_3\le\cdots.
\label{eq:spectrum}
\end{equation}
The intrinsic spectral time is
\begin{equation}
\tau_{\rm spec}=\frac{1}{\lambda_2}.
\label{eq:tauspec}
\end{equation}
Since both preparations evolve with the same $W(T_b)$,
\begin{equation}
\tau_{\rm spec}^H=\tau_{\rm spec}^C.
\label{eq:tauspec_equal}
\end{equation}
Thus a Markovian Mpemba effect does not require the hotter system to have a
different generator or spectral gap after the quench.

\subsection{Common-threshold relaxation time}

For a fixed absolute thermodynamic tolerance $\varepsilon>0$, define
\begin{equation}
\tau_{\rm th}^{\alpha}(\varepsilon)=\inf\{t\ge0:\,D_\alpha(t)\le\varepsilon\},
\qquad\alpha=H,C.
\label{eq:tauth}
\end{equation}
If $\varepsilon\ge D_\alpha(0)$ this time is zero. Logarithmic threshold
formulas below assume $0<\varepsilon<D_\alpha(0)$.
A direct operational Mpemba inequality is then
\begin{equation}
\tau_{\rm th}^H(\varepsilon)<\tau_{\rm th}^C(\varepsilon)
\quad\text{despite}\quad D_H(0)>D_C(0).
\label{eq:mpemba_ineq}
\end{equation}
The same threshold is used for both copies, so the comparison answers the
question: Which preparation first reaches the same neighborhood of the final
equilibrium? Equation~\eqref{eq:tauth} is the first-passage time of the
deterministic trajectory $D_\alpha(t)$ through the level $\varepsilon$.
When $D$ is estimated from finite samples, the estimated crossing time
fluctuates. Its distribution requires a specified observation or sampling
model and can then be studied with appropriate first-passage and
event-timing methods
\cite{Feller1971,CoxMiller1965,Kingman1993,Duki2016FirstPassageTime,Duki2018StochasticResonanceFirst2,Asfaw2010ThermallyActivatedBarrier,Asfaw2014ThermallyActivatedBarrier,Asfaw2012ExploringDynamicsDimer,Asfaw2010StochasticResonanceFlexible,Asfaw2011NoiseCreatedBistability,Birhanu2021StochasticResonatorLayered,Asfaw2013TimingStatisticsSpontaneous,Asfaw2013CorrectionTimingStatistics,Chen2012StatisticsCalciumMediated,Shiferaw2012CalciumMediatedTriggered}.

\subsection{Integrated thermodynamic relaxation time}

A threshold-free measure of the whole trajectory is
\begin{equation}
\bar\tau_{\alpha}=\frac{1}{D_\alpha(0)}\int_0^{\infty}D_\alpha(t)\,\mathrm dt.
\label{eq:taubar}
\end{equation}
This quantity measures the area under the normalized excess-free-energy curve.
Unlike Eq.~\eqref{eq:tauth}, it is sensitive to the entire relaxation history
rather than the time at which a single tolerance is crossed.

\subsection{Instantaneous thermodynamic relaxation time}

Equation~\eqref{eq:identity} suggests the local time scale
\begin{equation}
\tau_{\rm inst}(t)=\frac{D(t)}{\dot e_p(t)}
\label{eq:tauinst}
\end{equation}
whenever $D>0$ and $\dot e_p>0$.  Its reciprocal
\begin{equation}
r(t)=\frac{\dot e_p(t)}{D(t)}=-\frac{\mathrm d}{\mathrm dt}\ln D(t)
\label{eq:rdef}
\end{equation}
will be called the normalized entropy-production rate or thermodynamic
relaxation rate.  It quantifies the fraction of the remaining thermodynamic
distance dissipated per unit time.

The four time scales answer different questions.  $\tau_{\rm spec}$
characterizes the generator, $\tau_{\rm th}$ measures first arrival at a
prescribed thermodynamic accuracy, $\bar\tau$ measures the average persistence
of nonequilibrium free energy, and $\tau_{\rm inst}$ is a local rate-conversion
time.  Only the first is necessarily the same for hot and cold preparations.

\section{Exact entropy-production clock and Mpemba crossing}
\label{sec:clock}

\subsection{Thermodynamic clock representation}

Integrating Eq.~\eqref{eq:rdef} gives the exact identity
\begin{equation}
D_\alpha(t)=D_\alpha(0)\,\exp\!\left[-\Theta_\alpha(t)\right],
\qquad
\Theta_\alpha(t)=\int_0^tr_\alpha(s)\,\mathrm ds.
\label{eq:clock}
\end{equation}
In particular,
\begin{equation}
\Theta_\alpha(t)=\ln\frac{D_\alpha(0)}{D_\alpha(t)}.
\label{eq:clockmeaning}
\end{equation}
This dimensionless quantity counts the logarithmic reduction of the remaining
excess free energy: $\Theta=1$ means that $D$ has fallen to $D(0)/e$, and
$\Theta=\ln 10$ means a tenfold reduction. It contains the same information
as the normalized curve $D(t)/D(0)$. The word ``clock'' describes this
progress variable; it is not an additional physical time coordinate or a
new dynamical law. Both preparations share the same laboratory time.

Taking the ratio of the hot and cold thermodynamic distances yields
\begin{equation}
\ln\frac{D_H(t)}{D_C(t)}
=\ln\frac{D_H(0)}{D_C(0)}
-\int_0^t\left[r_H(s)-r_C(s)\right]\mathrm ds.
\label{eq:ratio}
\end{equation}
Let $t_M$ be the first positive time at which $D_H(t_M)=D_C(t_M)$,
assuming $D_H(0)>D_C(0)$. Such a meeting occurs precisely when
\begin{equation}
\int_0^{t_M}\!\left[\frac{\dot e_p^{H}(s)}{D_H(s)}-\frac{\dot e_p^{C}(s)}{D_C(s)}\right]
\mathrm ds=\ln\frac{D_H(0)}{D_C(0)}.
\label{eq:crossing}
\end{equation}
This is an exact condition, independent of a spectral approximation.
It identifies equality; a change of ordering additionally requires the
sign to change. The example below does exhibit this change. The bounds on
$t_M$ require only the first meeting, and do not assume that all possible
models have a unique crossing.

Equation~\eqref{eq:crossing} gives a simple interpretation.  The right-hand
side is the hot preparation's initial logarithmic free-energy disadvantage.
The left-hand side is its accumulated advantage in normalized entropy
production.  The hot preparation becomes closer to equilibrium only after the
latter has paid off the former.

It is useful to define the Mpemba relaxation balance
\begin{equation}
\mathcal M(t)=\int_0^t\left[r_H(s)-r_C(s)\right]\mathrm ds
-\ln\frac{D_H(0)}{D_C(0)}.
\label{eq:balance}
\end{equation}
Then
\begin{align}
\mathcal M(t)<0 &\;\Leftrightarrow\; D_H(t)>D_C(t),\nonumber\\
\mathcal M(t)=0 &\;\Leftrightarrow\; D_H(t)=D_C(t),
\label{eq:signs1}\\
\mathcal M(t)>0 &\;\Leftrightarrow\; D_H(t)<D_C(t).
\label{eq:signs2}
\end{align}

\subsection{A no-crossing criterion}

A direct corollary is a simple no-go statement.  If
\begin{equation}
r_H(t)\le r_C(t)
\label{eq:nogo}
\end{equation}
for all times before a putative crossing, while $D_H(0)>D_C(0)$, then
Eq.~\eqref{eq:ratio} cannot change sign and no Mpemba crossing is possible.
Thus a meeting requires the accumulated normalized-rate advantage to
reach the initial logarithmic deficit. Intervals with $r_H>r_C$ contribute
positively to that balance, while intervals with $r_H<r_C$ subtract from it;
a rate advantage at a single instant does not establish a crossing.

\subsection{Threshold times from the thermodynamic clock}

The threshold time in Eq.~\eqref{eq:tauth} satisfies the exact implicit
equation
\begin{equation}
\int_0^{\tau_{\rm th}^{\alpha}(\varepsilon)}r_\alpha(s)\,\mathrm ds
=\ln\frac{D_\alpha(0)}{\varepsilon}.
\label{eq:tauth_implicit}
\end{equation}
The hot preparation therefore has two competing features: a larger right-hand
side because it starts farther away, but potentially a larger accumulated
relaxation rate on the left.  The Mpemba effect occurs when the second
advantage wins.

\section{Relaxation modes of the common generator}
\label{sec:spectral}

The exact relations above require no linearization. To connect them with
the familiar slow-mode picture, we now decompose each trajectory into the
relaxation modes of the common generator $W$. This is the step that
explains why two preparations governed by the same dynamics can relax so
differently.

\subsection{Slow and fast relaxation patterns}

Because $W\pi_b=0$, the deviation from equilibrium,
$\delta p(t)=p(t)-\pi_b$, obeys the same equation as $p$ itself,
$\delta\dot p=W\,\delta p$. Let $v_n$ denote the eigenvectors of $W$,
\begin{equation}
Wv_n=-\lambda_n v_n,\qquad n=2,\dots,N,
\label{eq:modes}
\end{equation}
with the rates ordered as in Eq.~\eqref{eq:spectrum}. The stationary
vector $\pi_b$ is the eigenvector with $\lambda_1=0$, and since
probability is conserved, $\delta p$ has no component along it. Hence
\begin{equation}
p(t)=\pi_b+\sum_{n=2}^{N}b_n\,e^{-\lambda_nt}\,v_n .
\label{eq:pexp}
\end{equation}
Equation~\eqref{eq:pexp} contains the whole mechanism. The vector $v_2$ is
a slowest relaxation pattern, $v_3,\dots,v_N$ decay at least as fast, and the
amplitude $b_n$ measures how strongly the initial preparation excites
pattern $n$. The rates $\lambda_n$ belong to $W$ and are therefore common
to the hot and cold copies; only the amplitudes $b_n$ depend on the
initial temperature. For the three-state model,
$p(t)=\pi_b+b_2e^{-\lambda_2t}v_2+b_3e^{-\lambda_3t}v_3$.

Detailed balance makes the amplitudes simple to compute. The eigenvectors
of a detailed-balance generator are orthogonal once each component is
weighted by $1/\pi_{b,i}$ (Appendix~\ref{app:selfadjoint}), and we
normalize them accordingly:
\begin{equation}
\sum_i\frac{v_{m,i}\,v_{n,i}}{\pi_{b,i}}=\delta_{mn},
\qquad
b_n=\sum_i\frac{[p_i(0)-\pi_{b,i}]\,v_{n,i}}{\pi_{b,i}}.
\label{eq:orth}
\end{equation}

It is convenient to express the same expansion through the relative
deviation from equilibrium,
\begin{equation}
f_i(t)=\frac{p_i(t)}{\pi_{b,i}}-1
=\sum_{n=2}^{N}b_n\,e^{-\lambda_nt}\,\phi_{n,i},
\qquad
\phi_{n,i}=\frac{v_{n,i}}{\pi_{b,i}} .
\label{eq:fexp}
\end{equation}
Writing $\langle g\rangle_{\pi_b}=\sum_i\pi_{b,i}g_i$ for an equilibrium
average, Eq.~\eqref{eq:orth} reads
$\langle\phi_m\phi_n\rangle_{\pi_b}=\delta_{mn}$. The relative deviation
$f$ is the natural variable for the entropy: it measures, state by state,
the fractional excess or deficit of probability, and it will be used
again for the finite-amplitude bounds of Sec.~\ref{sec:envelopes}.

\subsection{Relative entropy and entropy production near equilibrium}

Expanding Eq.~\eqref{eq:Ddef} for $|f_i|\ll1$ gives
\begin{equation}
D(t)=\frac12\sum_i\frac{[\delta p_i(t)]^2}{\pi_{b,i}}+O(f^3)
=\frac12\sum_i\pi_{b,i}f_i(t)^2+O(f^3).
\label{eq:Dquad}
\end{equation}
Using orthonormality,
\begin{equation}
D(t)=\frac12\sum_{n=2}^Nb_n^2\,e^{-2\lambda_nt}+O(f^3).
\label{eq:Dmodes}
\end{equation}
Differentiation and Eq.~\eqref{eq:identity} give
\begin{equation}
\dot e_p(t)=\sum_{n=2}^N\lambda_nb_n^2\,e^{-2\lambda_nt}+O(f^3).
\label{eq:sigmamodes}
\end{equation}
Consequently,
\begin{equation}
r(t)=2\,\frac{\sum_{n\ge2}\lambda_nb_n^2e^{-2\lambda_nt}}
{\sum_{n\ge2}b_n^2e^{-2\lambda_nt}}+O(f).
\label{eq:rmodes}
\end{equation}
Thus the normalized entropy-production rate is twice a time-dependent
mode-weighted average of the relaxation eigenvalues.  In particular,
\begin{equation}
2\lambda_2\le r_{\rm quad}(t)\le2\lambda_N,
\qquad r(t)=r_{\rm quad}(t)+O(\|f(t)\|_\infty),
\label{eq:rbounds}
\end{equation}
and
\begin{equation}
\frac{1}{2\lambda_N}\le\frac{1}{r_{\rm quad}(t)}\le\frac{1}{2\lambda_2}.
\label{eq:tauinstbounds}
\end{equation}
We emphasize that Eqs.~\eqref{eq:rbounds}--\eqref{eq:tauinstbounds} hold only
to quadratic order in the deviation from equilibrium.  They are \emph{not}
global statements: as shown in Sec.~\ref{sec:bounds}, the exact rate can
undershoot $2\lambda_2$ through cubic entropy corrections when the slow
eigenfunction has a nonzero third moment. The correct uniform floor is set by
the modified logarithmic Sobolev constant of the generator.

The integrated thermodynamic time also has a transparent mode representation.
From Eq.~\eqref{eq:Dmodes},
\begin{equation}
\bar\tau=\frac{\sum_{n\ge2}b_n^2/(2\lambda_n)}{\sum_{n\ge2}b_n^2}+O(f).
\label{eq:taubarmodes}
\end{equation}
Writing
\begin{equation}
w_n=\frac{b_n^2}{\sum_{m\ge2}b_m^2},\qquad\sum_{n\ge2}w_n=1,
\label{eq:weights}
\end{equation}
we obtain
\begin{equation}
\bar\tau=\sum_{n\ge2}\frac{w_n}{2\lambda_n}+O(f).
\label{eq:taubarweighted}
\end{equation}
The integrated relaxation time is therefore preparation dependent even though
the spectrum is common.

\subsection{Asymptotic threshold time}

Let $m$ be the smallest index such that $b_m\ne0$, and define
$\mathcal B_m^2=\sum_{n:\lambda_n=\lambda_m}b_n^2$ to include any
degeneracy of that active eigenvalue. At long times,
\begin{equation}
D(t)\simeq\frac{\mathcal B_m^2}{2}\,e^{-2\lambda_mt}.
\label{eq:Dasymp}
\end{equation}
For sufficiently small $\varepsilon$, the threshold time is
\begin{equation}
\tau_{\rm th}(\varepsilon)\simeq\frac{1}{2\lambda_m}
\ln\!\left(\frac{\mathcal B_m^2}{2\varepsilon}\right).
\label{eq:tauth_asymp}
\end{equation}
This formula makes clear that an operational relaxation time depends on both a
decay rate and an initial amplitude.

\section{Strong Mpemba effect as a change of thermodynamic time scale}
\label{sec:strong}

Assume the slow eigenvalue is simple, as in our example. In the generic
case $b_2\ne0$, the slowest nonstationary mode eventually
dominates,
\begin{equation}
D(t)\sim\frac{b_2^2}{2}\,e^{-2\lambda_2t},
\qquad
\tau_{\rm inst}(t)\to\frac{1}{2\lambda_2}.
\label{eq:generic}
\end{equation}
A strong Mpemba preparation is a temperature $T_\star>T_b$ at which the
coefficient of the slowest mode vanishes,
\begin{equation}
b_2(T_\star)=0.
\label{eq:strongdef}
\end{equation}
For a three-state system the next mode is $n=3$, so
\begin{equation}
D_\star(t)\sim\frac{b_3(T_\star)^2}{2}\,e^{-2\lambda_3t},
\qquad
\tau_{\rm inst}^\star(t)\to\frac{1}{2\lambda_3}.
\label{eq:strongdecay}
\end{equation}
The final generator still has the same spectral gap $\lambda_2$, but that mode
is absent from the selected initial condition.  Consequently the thermodynamic
distance decays on the faster scale $1/(2\lambda_3)$.

This distinction is worth emphasizing: the two preparations share the
same generator and the same spectral time $\tau_{\rm spec}=1/\lambda_2$,
but they have different preparation-dependent thermodynamic relaxation
times.
The strong Mpemba effect does not alter the spectrum. It alters which parts of
that spectrum are populated by the initial canonical state. If the slow
eigenvalue is degenerate, its entire eigenspace must be canceled.

Exact cancellation is an ideal preparation condition. With a small residual
$b_2\ne0$, the slow tail eventually returns. Comparing the two quadratic
mode contributions gives the crossover scale
\begin{equation}
t_{\rm tail}\simeq\frac{\ln|b_3/b_2|}{\lambda_3-\lambda_2}
\quad\text{when }|b_3|>|b_2|.
\label{eq:tuning}
\end{equation}
The strong-preparation limits below refer to the exact root $T_\star$;
finite-precision roots and experimental temperature errors approximate it.

\section{Global entropy-decay and crossing constraints}
\label{sec:bounds}

The inequalities in this section apply beyond the quadratic approximation.
Their entropy-decay ingredient is the established modified logarithmic
Sobolev inequality \cite{bobkov2006}; our use of it concerns the crossing of
two preparations. Close thermodynamic connections are discussed in
Refs.~\cite{bao2025,auconi2026}.

\subsection{The entropy--entropy-production constant}

For a fixed reversible generator define
\begin{equation}
\rho=\inf_{p\ne\pi_b}\frac{\dot e_p(p)}{2D[p\Vert\pi_b]},
\label{eq:rhodef}
\end{equation}
with the infimum over strictly positive probability distributions.
Writing $h=p/\pi_b$, the Dirichlet form is
\begin{equation}
\mathcal E(h,\ln h)
=\frac12\sum_{i,j}\pi_{b,i}k_{ij}(h_i-h_j)(\ln h_i-\ln h_j)
=\dot e_p(p).
\label{eq:dirichlet}
\end{equation}
Thus $\rho D\le\mathcal E(h,\ln h)/2$. This agrees with the constant
$\rho_0$ in Eq.~(1.5) of Bobkov and Tetali \cite{bobkov2006}; conventions
elsewhere can differ by a factor of two. Classical logarithmic Sobolev
constants involve a different functional \cite{diaconis1996}; they should
not be equated with the modified constant without comparing definitions. In a finite irreducible chain,
$\rho>0$.

\emph{Proposition 1 (standard entropy-decay inequality).}
For a nonstationary trajectory,
\begin{equation}
r(t)\ge2\rho,\qquad D(t)\le D(0)e^{-2\rho t},
\qquad \tau_{\rm inst}(t)\le\frac1{2\rho}.
\label{eq:rfloor}
\end{equation}
Consequently,
\begin{equation}
\bar\tau\le\frac1{2\rho},\qquad
\tau_{\rm th}(\varepsilon)\le
\frac{[\ln(D(0)/\varepsilon)]_+}{2\rho},
\quad [x]_+=\max(x,0).
\label{eq:expdecay}
\end{equation}
At $p=\pi_b$, $D=\dot e_p=0$ and their ratio is undefined; the decay
inequality still holds. Boundary initial states can be handled by taking
$t>0$ and a limit. Appendix~\ref{app:bounds} gives the short proof.

It is essential to distinguish an estimate of the infimum from a lower
bound on it. Evaluating $\dot e_p/(2D)$ at any trial distribution gives an
\emph{upper} bound on $\rho$. Local minimization, even with many restarts,
does not reverse that direction. We write $\widehat\rho$ for the numerical
estimate and $\underline\rho$ for a proved lower bound. Replacing $\rho$
by $\underline\rho$ in Eqs.~\eqref{eq:rfloor}--\eqref{eq:expdecay}
preserves their validity. In a fully connected chain one convenient
certificate is
\begin{equation}
\kappa=\min_{i\ne j}\frac{k_{ij}}{\pi_{b,j}},\qquad
\underline\rho=\frac\kappa2\le\rho.
\label{eq:resetbound}
\end{equation}
It follows by subtracting a reset generator with rates $\kappa\pi_{b,j}$;
the residual generator still dissipates relative entropy. The explicit
derivation is in Appendix~\ref{app:reset}. This elementary certificate is
useful for the present complete graph, but can be zero on a sparse graph.

\subsection{When is the spectral floor strictly too large?}

Approaching equilibrium along a normalized slow eigenfunction gives
$\rho\le\lambda_2$. To establish strictness, take
$p_\epsilon=\pi_b(1+\epsilon\phi_2)$, with
$\langle\phi_2^2\rangle_{\pi_b}=1$ and
$\mu_3=\langle\phi_2^3\rangle_{\pi_b}$. Then
\begin{equation}
\frac{\dot e_p(p_\epsilon)}{2D[p_\epsilon\Vert\pi_b]}
=\lambda_2\left[1-\frac{\mu_3}{6}\epsilon+O(\epsilon^2)\right].
\label{eq:cubiccriterion}
\end{equation}
If $\mu_3\ne0$, choosing $\epsilon$ with the same sign as $\mu_3$
proves $\rho<\lambda_2$. If $\mu_3=0$, this argument decides nothing:
higher-order terms or other directions must be examined. A symmetric
two-state chain has $\rho=\lambda_2$, so strict inequality is not universal.
Appendix~\ref{app:cubic} derives the expansion and the counterexample.

\subsection{A trajectory upper rate is also needed}
\label{sec:noupper}

No finite constant bounds $\dot e_p/(2D)$ from above over the entire
probability simplex. Sending an occupation toward zero while a nonzero
influx remains makes a logarithmic term in $\dot e_p$ diverge, while $D$
stays finite. This concerns an instantaneous fractional rate; it does not
by itself prove an unbounded speedup at a fixed threshold.
For an interior initial condition define the finite trajectory quantity
\begin{equation}
R_H=\sup_{t\ge0}r_H(t).
\label{eq:Rdef}
\end{equation}
A truncated numerical scan is a lower estimate of this supremum, not a
valid upper bound. We use a proved upper bound $\overline R_H\ge R_H$.
For the exact strong preparation in our model,
Appendix~\ref{app:hotcap} proves $R_H=2\lambda_3$, approached as
$t\to\infty$ rather than attained at a finite scan endpoint.

\subsection{A global crossing-time constraint}

\emph{Proposition 2 (rate-based crossing bound).}
If $D_H(0)>D_C(0)$ and a meeting occurs, then
\begin{equation}
t_M\ge\frac{\ln[D_H(0)/D_C(0)]}{R_H-2\rho}
\ge\frac{\ln[D_H(0)/D_C(0)]}
{\overline R_H-2\underline\rho}.
\label{eq:tMbound}
\end{equation}
The denominators are positive whenever a meeting with this initial ordering
is possible. Indeed Eq.~\eqref{eq:crossing} gives
\begin{align}
\ln\frac{D_H(0)}{D_C(0)}
&=\int_0^{t_M}(r_H-r_C)\,\mathrm ds\nonumber\\
&\le t_M(R_H-2\rho)
\le t_M(\overline R_H-2\underline\rho).
\label{eq:rateproof}
\end{align}
This is an elementary consequence of valid rate bounds, not an independent
universal speed-limit principle. For any known $s<t_M$, the same proof
restarts with $\ln[D_H(s)/D_C(s)]$ and bounds valid for $t\ge s$.
The next section improves the constraint by retaining the actual mode
amplitudes instead of only the largest and smallest possible rates.

\section{Crossing bounds that retain the mode amplitudes}
\label{sec:envelopes}

\subsection{Exact entropy envelopes at finite amplitude}

The static comparison between relative entropy and
$Q=\chi^2(p\Vert\pi_b)$ is an established bounded-likelihood-ratio
inequality \cite{sasonverdu2016}: for natural logarithms,
$F(\max_ip_i/\pi_{b,i}-1)\le D(p\Vert\pi_b)/\chi^2(p\Vert\pi_b)
\le F(\min_ip_i/\pi_{b,i}-1)$, with $F$ defined in
Eq.~\eqref{eq:Fscalar} below. Here we combine that comparison with the
Markov maximum principle and the preparation-dependent mode amplitudes to
construct iterative bounds on the first relative-entropy crossing. We
include the scalar derivation to keep the application self-contained.

Define, for $x>-1$, the scalar functions
\begin{equation}
g(x)=(1+x)\ln(1+x)-x,\quad
F(x)=\begin{cases}g(x)/x^2,&x\ne0,\\[2pt]1/2,&x=0,\end{cases}
\label{eq:Fscalar}
\end{equation}
Since $g(0)=g'(0)=0$ and $g''(x)=1/(1+x)$, integrating twice gives
\begin{equation}
\begin{gathered}
F(x)=\int_0^1\frac{1-y}{1+yx}\,\mathrm dy,\\
F'(x)=-\int_0^1\frac{y(1-y)}{(1+yx)^2}\,\mathrm dy<0.
\end{gathered}
\label{eq:Fdecrease}
\end{equation}
Normalization removes the linear term in the entropy, so
$D=\sum_i\pi_{b,i}g(f_i)$ exactly. For a reference time $s$, let
\begin{equation}
\begin{gathered}
m_\alpha(s)=\min_i f_i^\alpha(s),\qquad
M_\alpha(s)=\max_i f_i^\alpha(s),\\
Q_\alpha(t)=\sum_i\pi_{b,i}[f_i^\alpha(t)]^2.
\end{gathered}
\label{eq:envelopedefs}
\end{equation}
From $p(t)=e^{W(t-s)}p(s)$ and $W\pi_b=0$,
each component evolves as $f_i(t)=\sum_jK_{ij}f_j(s)$ with weights
$K_{ij}=[e^{W(t-s)}]_{ij}\pi_{b,j}/\pi_{b,i}\ge0$ that satisfy
$\sum_jK_{ij}=1$. Each $f_i(t)$ is therefore a weighted average of the
values at time $s$, and every component at $t\ge s$ remains
between $m_\alpha(s)$ and $M_\alpha(s)$. Monotonicity of $F$ gives the
finite-amplitude bounds
\begin{equation}
F(M_\alpha(s))Q_\alpha(t)
\le D_\alpha(t)\le
F(m_\alpha(s))Q_\alpha(t),\qquad t\ge s.
\label{eq:envelopes}
\end{equation}
Unlike $D\simeq Q/2$, these inequalities have no omitted Taylor remainder
and require no small-deviation assumption. In the notation of probability
divergences, $Q_\alpha(t)=\chi^2(p^\alpha(t)\Vert\pi_b)$.
As the reference state $p^\alpha(s)$ approaches equilibrium, both factors
tend to $1/2$. For a fixed reference time $s$ the factors remain fixed;
updating $s$ is the step that tightens the dynamical enclosure.

\subsection{A three-state crossing interval}

Suppose the hot preparation is exactly strong, the cold preparation has a
nonzero slow mode, and $\lambda_3>\lambda_2>0$. Set
\begin{equation}
A=b_{3,H}^2,\quad B=b_{2,C}^2>0,\quad C=b_{3,C}^2,
\quad\Delta=\lambda_3-\lambda_2.
\label{eq:ABC}
\end{equation}
The exact squared amplitudes are
\begin{equation}
Q_H(t)=Ae^{-2\lambda_3t},\qquad
Q_C(t)=Be^{-2\lambda_2t}+Ce^{-2\lambda_3t}.
\label{eq:Qmodes}
\end{equation}
Abbreviate $a_\alpha(s)=F(M_\alpha(s))$ and
$c_\alpha(s)=F(m_\alpha(s))$. Whenever the logarithms below have positive
arguments, define
\begin{align}
\ell(s)&=\frac1{2\Delta}
\ln\!\left[\frac{a_H(s)A-c_C(s)C}{c_C(s)B}\right],
\label{eq:loweramp}\\
u(s)&=\frac1{2\Delta}
\ln\!\left[\frac{c_H(s)A-a_C(s)C}{a_C(s)B}\right].
\label{eq:upperamp}
\end{align}
For a known lower bound $s\le t_M$, comparison of the lower hot envelope
with the upper cold envelope implies
\begin{equation}
t_M\ge\max\{s,\ell(s)\}.
\label{eq:amp_lower}
\end{equation}
Comparison of the upper hot envelope with the lower cold envelope ensures
$D_H(t)\le D_C(t)$ for $t\ge\max\{s,u(s)\}$, and continuity then gives
\begin{equation}
t_M\le\max\{s,u(s)\}.
\label{eq:amp_upper}
\end{equation}
These statements bound the first meeting without requiring uniqueness.
If the numerator in Eq.~\eqref{eq:loweramp} is nonpositive, there is no
additional lower constraint. If the numerator in Eq.~\eqref{eq:upperamp}
is nonpositive, the upper-envelope comparison already holds at $s$.

Start from $s_0=0$ and successively set
$s_{n+1}=\max\{s_n,\ell(s_n)\}$. Every iterate remains a proved lower
bound. Re-evaluating the envelopes at these later times tightens them because
the component range has contracted. Only the generator, initial amplitudes,
and distributions at known reference times are needed; no global
minimization of $\dot e_p/(2D)$ or numerical solution of the crossing equation
is used in constructing the bounds. Appendix~\ref{app:envelopes} supplies
the comparison steps explicitly. The evaluated bounds are reported after
the model is specified.

\section{Explicit three-state detailed-balance model}
\label{sec:model}

\subsection{Energies and rates}

We now construct a minimal example with three states.  Energies are measured
in units of $T_b$ and time in units of the inverse reference attempt rate.  We set
\begin{equation}
T_b=1,\qquad (E_0,E_1,E_2)=(0,\,0.4,\,3.0).
\label{eq:energies}
\end{equation}
For each undirected edge choose a symmetric kinetic prefactor
$\gamma_{ij}=\gamma_{ji}$ and rates
\begin{equation}
k_{ij}=\gamma_{ij}\exp\!\left[-\frac{E_j-E_i}{2T_b}\right].
\label{eq:rates}
\end{equation}
Equation~\eqref{eq:rates} guarantees
\begin{equation}
\frac{k_{ij}}{k_{ji}}=e^{-(E_j-E_i)/T_b},
\label{eq:ratio_rates}
\end{equation}
so detailed balance holds exactly.  We choose
\begin{equation}
\gamma_{01}=1,\qquad \gamma_{12}=0.12,\qquad \gamma_{02}=4.6 .
\label{eq:gammas}
\end{equation}
These unequal kinetic prefactors represent unequal transition barriers.  Here
kinetic asymmetry produces a nontrivial orientation of the
relaxation eigenmodes relative to the one-parameter family of canonical
initial states.

Using the column-vector convention $\dot p=Wp$, the resulting matrix is
\begin{equation}
W=\begin{pmatrix}
-1.84512949 & \phantom{-}1.22140276 & \phantom{-}20.61576972\\
\phantom{-}0.81873075 & -1.25410657 & \phantom{-}0.44031560\\
\phantom{-}1.02639874 & \phantom{-}0.03270382 & -21.05608532
\end{pmatrix},
\label{eq:Wmatrix}
\end{equation}
whose columns sum to zero.  The equilibrium distribution is
\begin{equation}
\pi_b=(0.58135914,\;0.38969669,\;0.02894417)^{\rm T}.
\label{eq:pib}
\end{equation}
The nonzero relaxation rates are
\begin{equation}
\lambda_2=2.05406255,\qquad \lambda_3=22.1012588 .
\label{eq:lambdas}
\end{equation}
Therefore
\begin{equation}
\tau_{\rm spec}=\frac{1}{\lambda_2}=0.4868401,
\label{eq:tauspec_num}
\end{equation}
while the near-equilibrium thermodynamic times associated with the individual
modes are
\begin{equation}
\frac{1}{2\lambda_2}=0.2434200,\qquad
\frac{1}{2\lambda_3}=0.0226231 .
\label{eq:modetimes}
\end{equation}

\begin{figure}[!tbp]
\centering
\includegraphics[width=\columnwidth]{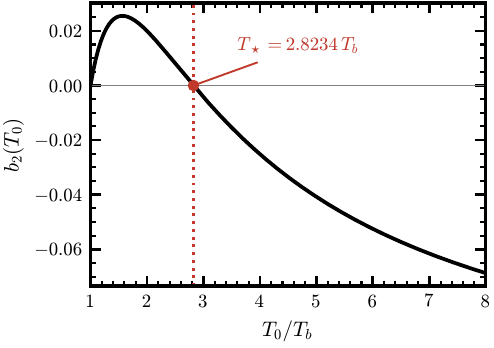}
\caption{Coefficient of the slowest nonstationary mode as a function of the
initial canonical temperature for the three-state model.  The zero at
$T_\star\simeq2.82343\,T_b$ identifies a strong-Mpemba preparation: the final
generator is unchanged, but the initial state has no projection onto its
slowest mode.}
\label{fig:a2}
\end{figure}

\subsection{Strong-Mpemba temperature}

Expanding the canonical initial state $\pi(T_0)$ in the eigenbasis of $W$, the
normalized slow-mode coefficient $b_2(T_0)$ crosses zero at
\begin{equation}
T_\star=2.8234286567\,T_b .
\label{eq:Tstar}
\end{equation}
Figure~\ref{fig:a2} shows this zero explicitly. The displayed decimal is a
numerical approximation to the root $b_2(T_\star)=0$, which defines the
exact ideal preparation. The mode with rate $\lambda_2$ is then absent,
so the asymptotic relative entropy is controlled by $\lambda_3$.

\section{Numerical relaxation of hot and cold preparations}
\label{sec:numerics}

We compare the strong-Mpemba hot preparation
\begin{equation}
T_H=T_\star=2.8234286567\,T_b
\label{eq:THnum}
\end{equation}
with a colder preparation
\begin{equation}
T_C=1.5\,T_b .
\label{eq:TCnum}
\end{equation}
Both are propagated by the same matrix in Eq.~\eqref{eq:Wmatrix}; no kinetic
parameter is changed between the two runs.

\begin{figure}[!tbp]
\centering
\includegraphics[width=\columnwidth]{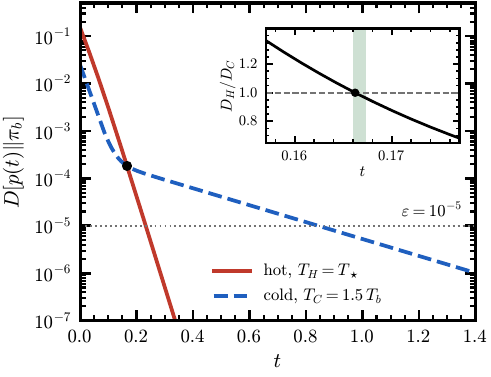}
\caption{Relative entropy to the common final equilibrium state for the hot
strong-Mpemba preparation and the colder preparation.  Although
$D_H(0)>D_C(0)$, the curves cross at $t_M\simeq0.1662$.  The inset resolves the ratio $D_H/D_C$ near the crossing; the shaded band
is the amplitude-based crossing interval. The horizontal dotted line indicates the common threshold $\varepsilon=10^{-5}$ used to define the
operational relaxation times.}
\label{fig:D}
\end{figure}

\subsection{Thermodynamic distance and crossing}

The initial relative entropies are
\begin{equation}
D_H(0)=0.1515934,\qquad D_C(0)=0.0247632,
\label{eq:D0s}
\end{equation}
so the hotter state begins approximately
\begin{equation}
\frac{D_H(0)}{D_C(0)}=6.12
\label{eq:D0ratio}
\end{equation}
times farther from the final equilibrium state in thermodynamic distance.
Nevertheless, the two curves cross at
\begin{equation}
t_M=0.1662209 .
\label{eq:tMnum}
\end{equation}
After this time the initially hotter system is closer to equilibrium, as shown
in Fig.~\ref{fig:D}. Appendix~\ref{app:unique_crossing} proves that this
example has only one crossing and that the ordering persists afterwards.

\subsection{Mean energy is a different observable}

The model also shows why a relative-entropy crossing should not be called
a temperature or freezing crossing without further analysis. At the entropy
crossing, the mean energies are
\begin{equation}
U_H(t_M)\simeq0.2524200>U_C(t_M)\simeq0.2481862.
\label{eq:energy_at_KL}
\end{equation}
The hotter preparation is therefore still higher in mean energy when the
two relative entropies meet. Here the energy difference has the exact
two-mode form
\begin{equation}
U_H(t)-U_C(t)=a_Ee^{-\lambda_3t}-b_Ee^{-\lambda_2t},
\label{eq:energy_difference}
\end{equation}
with $a_E\simeq0.2536261$ and $b_E\simeq0.00310078$. Its crossing is
\begin{equation}
t_U=\frac{\ln(a_E/b_E)}{\lambda_3-\lambda_2}
\simeq0.219692>t_M.
\label{eq:energy_crossing}
\end{equation}
The coefficients follow by applying $U=\sum_i E_ip_i$ to the exact mode
expansion. The two observables probe different features of the same
relaxation and need not reverse their ordering simultaneously.

\subsection{Entropy production}

\begin{figure}[!tbp]
\centering
\includegraphics[width=\columnwidth]{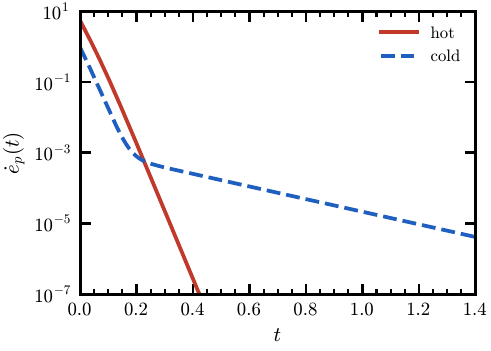}
\caption{Entropy-production rates for the same two preparations.  Detailed
balance ensures $\dot e_p(t)=-\mathrm dD/\mathrm dt$.  The initially hotter
state has a substantially larger early entropy-production rate, but the Mpemba
crossing is controlled by the normalized rate $\dot e_p/D$, not by $\dot e_p$
alone.}
\label{fig:sigma}
\end{figure}

The corresponding entropy-production rates are shown in Fig.~\ref{fig:sigma}.
At $t=0$,
\begin{equation}
\dot e_p^{H}(0)=5.45960,\qquad \dot e_p^{C}(0)=0.951551 .
\label{eq:sigma0}
\end{equation}
The hotter preparation initially dissipates free energy much more rapidly in
absolute terms.  Absolute entropy production alone, however, is not sufficient
to determine a crossing because the hotter state also contains more excess
free energy.  The relevant quantity for the crossing identity is the
normalized rate $\dot e_p/D$.

\begin{figure}[!tbp]
\centering
\includegraphics[width=\columnwidth]{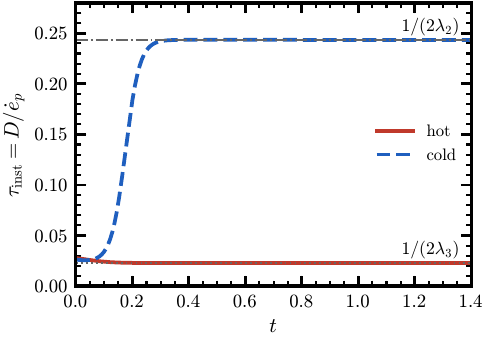}
\caption{Instantaneous thermodynamic relaxation time
$\tau_{\rm inst}=D/\dot e_p$.  The hot strong-Mpemba preparation approaches the
fast-mode value $1/(2\lambda_3)$, whereas the colder preparation approaches
$1/(2\lambda_2)$.  Both copies nevertheless have the same generator and the
same spectral time $1/\lambda_2$.}
\label{fig:tauinst}
\end{figure}

\subsection{Instantaneous thermodynamic relaxation time}

Figure~\ref{fig:tauinst} displays
\begin{equation}
\tau_{\rm inst}^{\alpha}(t)=\frac{D_\alpha(t)}{\dot e_p^{\alpha}(t)} .
\label{eq:tauinst_num}
\end{equation}
At very early times the two local time scales are comparable. The ideal
strong-Mpemba probability deviation evolves entirely along the fast mode
from $t=0$. As its amplitude decreases, the nonlinear entropy-to-production
ratio approaches
\begin{equation}
\tau_{\rm inst,H}\longrightarrow\frac{1}{2\lambda_3}=0.0226231,
\label{eq:tauinstH}
\end{equation}
whereas the colder preparation retains a nonzero slow-mode component and
approaches
\begin{equation}
\tau_{\rm inst,C}\longrightarrow\frac{1}{2\lambda_2}=0.2434200 .
\label{eq:tauinstC}
\end{equation}
The order-of-magnitude difference is a direct thermodynamic signature of
slow-mode cancellation.

\subsection{Threshold and integrated relaxation times}

For a common thermodynamic tolerance
\begin{equation}
\varepsilon=10^{-5},
\label{eq:eps}
\end{equation}
the threshold times are
\begin{equation}
\tau_{\rm th}^H=0.232571,\qquad
\tau_{\rm th}^C=0.843811 .
\label{eq:tauth_num}
\end{equation}
Thus
\begin{equation}
\frac{\tau_{\rm th}^C}{\tau_{\rm th}^H}=3.63 .
\label{eq:tauth_ratio}
\end{equation}
The hotter state reaches the same absolute thermodynamic tolerance in
about $27.6\%$ of the cold preparation's time, despite beginning much farther away.

The integrated times are
\begin{equation}
\bar\tau^{\,H}=0.0266964,\qquad
\bar\tau^{\,C}=0.0276949 .
\label{eq:taubar_num}
\end{equation}
The contrast is smaller because $\bar\tau$ weights the entire early
trajectory, where both states initially contain appreciable fast components.
This difference between metrics is physically informative rather than
contradictory: the threshold time is highly sensitive to the asymptotic slow
tail, whereas the integrated time gives substantial weight to early-time
relaxation.

\begin{figure*}[!t]
\centering
\includegraphics[width=\textwidth]{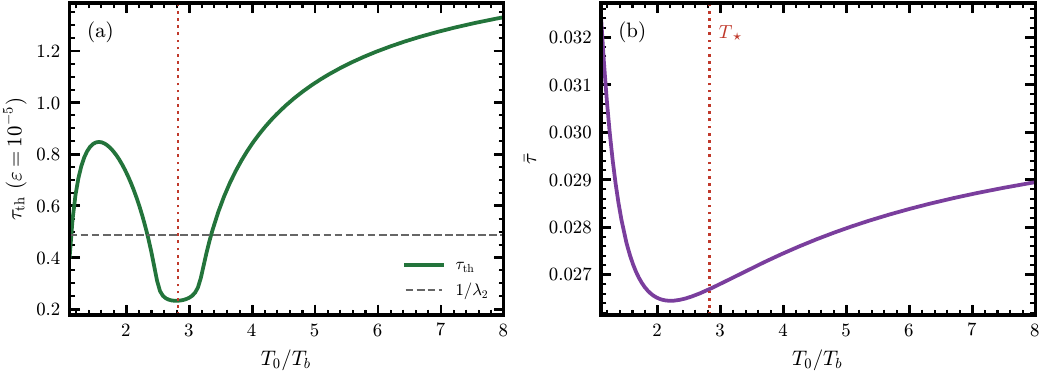}
\caption{Relaxation times versus initial temperature for the same final
generator. (a) The common-threshold time at $\varepsilon=10^{-5}$, with
the preparation-independent spectral time $1/\lambda_2$ as a reference.
(b) The integrated thermodynamic time, shown on a separate vertical scale
to resolve its weaker variation. The vertical line marks $T_\star$.}
\label{fig:times}
\end{figure*}

Figure~\ref{fig:times} scans both quantities over the initial temperature.
The spectral time is constant because it is a property of the final generator.
The preparation-dependent times vary nonmonotonically; the threshold time
shows a sharp minimum near the strong-Mpemba temperature, whereas the
integrated time has a much weaker variation. The minimizing preparation
depends on the time measure and need not coincide exactly with $T_\star$.

\subsection{Estimates, certificates, and the refined crossing interval}
\label{sec:boundscheck}

Multistart minimization of $\dot e_p/(2D)$ gives
\begin{equation}
\begin{gathered}
\widehat\rho=2.0371504\simeq0.99177\lambda_2,\\
p_{\rm cand}\simeq(0.4766385,0.4998548,0.0235067).
\end{gathered}
\label{eq:rhovalue}
\end{equation}
This is a numerical estimate, not a certified global minimum. The independent
analytic certificate from Eq.~\eqref{eq:resetbound} is
\begin{equation}
\begin{gathered}
\kappa\simeq1.1298931,\qquad
\underline\rho=\kappa/2\simeq0.5649466,\\
\underline\rho\le\rho
\le\frac{\dot e_p(p_{\rm cand})}{2D(p_{\rm cand}\Vert\pi_b)}.
\end{gathered}
\label{eq:rhocertificate_num}
\end{equation}
The upper endpoint evaluates to Eq.~\eqref{eq:rhovalue}. Its large separation
from the conservative lower certificate is disclosed rather than hidden.
Independently, $|\mu_3|\simeq0.4522296\ne0$ for the normalized slow
eigenfunction, so Eq.~\eqref{eq:cubiccriterion} proves strict
$\rho<\lambda_2$ for this model without asserting that the optimizer is
global. The cold trajectory has
$\min_t r_C(t)/2\simeq2.052549<\lambda_2$.

For the ideal strong preparation, Appendix~\ref{app:hotcap} proves
$R_H=2\lambda_3\simeq44.2025177$. Combining this with the certified floor gives
\begin{equation}
t_M\ge\frac{\ln[D_H(0)/D_C(0)]}{2\lambda_3-\kappa}
\simeq0.0420648.
\label{eq:tMbound_num}
\end{equation}
Substituting $\widehat\rho$ instead would give about $0.04515$, but that
number is only an estimate of the optimized rate-based constraint and is not
used as a certified bound.

The mode envelopes are much sharper. Iterating
Eqs.~\eqref{eq:loweramp}--\eqref{eq:amp_upper} gives the evaluations in
Table~\ref{tab:envelope_iterations}. After a few steps the lower and upper
expressions are approximately $0.16595$ and $0.16739$. With outward rounding
at the displayed precision,
\begin{equation}
\boxed{0.1659\le t_M\le0.1674},
\label{eq:tMrefined_num}
\end{equation}
which brackets the root $t_M\simeq0.1662209$ of $D_H(t)=D_C(t)$.
The enclosure follows from analytic inequalities, and an independent
outward-rounded interval calculation certifies its displayed endpoints,
the strong-preparation root, and the sign conditions used in the proofs
(Appendix~\ref{app:interval}). The ordinary and 60-digit point evaluations
provide separate reproducibility checks; they are not the interval
certificate. The gap between the refined lower
expression and the numerical crossing is about $0.16\%$, compared with the
much looser global constraint. The improvement comes from retaining the
cold preparation's small slow-mode amplitude and both preparations' fast
amplitudes.

\begin{table}[t]
\centering
\caption{Successive reference times known to precede the first crossing.
Numbers shown are rounded evaluations of the analytic envelope formulas.
The interval-certified implementation uses downward-rounded lower endpoints
as its next references, as detailed in Appendix~\ref{app:interval}.}
\label{tab:envelope_iterations}
\begin{tabular}{rrrr}
\toprule
Step & Reference $s$ & Lower $\ell(s)$ & Upper $u(s)$\\
\midrule
0 & 0.000000 & 0.144456 & 0.178830\\
1 & 0.144456 & 0.165345 & 0.167623\\
2 & 0.165345 & 0.165936 & 0.167392\\
3 & 0.165936 & 0.165949 & 0.167387\\
4 & 0.165949 & 0.165950 & 0.167386\\
\bottomrule
\end{tabular}
\end{table}

\begin{figure}[!tbp]
\centering
\includegraphics[width=\columnwidth]{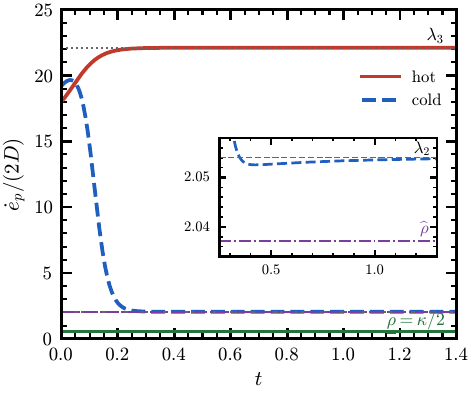}
\caption{Normalized entropy-production rate $r(t)/2$ for both preparations.
The analytic lower certificate $\underline\rho=\kappa/2$ is distinguished
from the estimated optimum $\widehat\rho$. The latter is not used as a
certified floor. The hot rate approaches $\lambda_3$ as $t\to\infty$.
Inset: the cold trajectory dips slightly below $\lambda_2$, consistent
with the sufficient cubic-moment condition.}
\label{fig:clockrate}
\end{figure}

Figure~\ref{fig:clockrate} distinguishes the proved lower certificate,
the numerical estimate, and the two spectral rates. It also resolves the
small cold-state dip below $\lambda_2$. Agreement of particular trajectories
with $\widehat\rho$ is a consistency check, not evidence of a global proof.

\section{Relaxation to a nonequilibrium steady state}
\label{sec:ness}

\subsection{The nonadiabatic entropy-production balance}

Let $W$ now be a fixed irreducible generator, with bidirectional allowed
transitions and stationary distribution $\pi_{\rm ness}$. Detailed balance
need not hold. The standard decomposition is
\cite{esposito2010,hatano2001}
\begin{equation}
\dot e_p^{\rm tot}=\dot e_p^{\rm na}+\dot e_p^{\rm hk},
\qquad \dot e_p^{\rm na}\ge0,\quad\dot e_p^{\rm hk}\ge0,
\label{eq:split}
\end{equation}
where
\begin{align}
\dot e_p^{\rm na}&=-\sum_i\dot p_i\ln\frac{p_i}{\pi_{{\rm ness},i}},
\label{eq:signa}\\
\dot e_p^{\rm hk}&=\sum_{i,j\ne i}p_i k_{ij}
\ln\frac{k_{ij}\pi_{{\rm ness},i}}{k_{ji}\pi_{{\rm ness},j}}.
\label{eq:sighk}
\end{align}
We use ``nonadiabatic'' as the precise term; terminology for excess heat and
excess entropy production is not identical across all conventions.
With the stationary reference fixed,
\begin{equation}
-\frac{\mathrm d}{\mathrm dt}D[p(t)\Vert\pi_{\rm ness}]
=\dot e_p^{\rm na}(t).
\label{eq:identity_ness}
\end{equation}
Appendix~\ref{app:ness} proves the identity and positivity explicitly.
This is established stationary-state thermodynamics, including in previous
relaxation and Mpemba analyses \cite{bao2025,summer2026}.

The clock and rate-based crossing relations hold with
\begin{equation}
r_{\rm na}=\frac{\dot e_p^{\rm na}}D,\qquad
\rho_{\rm na}=\inf_{p\ne\pi_{\rm ness}}
\frac{\dot e_p^{\rm na}(p)}{2D[p\Vert\pi_{\rm ness}]}.
\label{eq:replacements}
\end{equation}
The orthogonal spectral formulas in Sec.~\ref{sec:spectral} and the
three-state amplitude formulas in Sec.~\ref{sec:envelopes} require
reversibility and are not transferred unchanged to a general driven
generator. In particular the local dissipation form involves the symmetric
part of the generator in the stationary inner product, not generally its
slowest ordinary eigenvalue.

At a fixed generator, housekeeping dissipation is absent from $-\dot D$.
This does not mean that driving is dynamically irrelevant: changing the
drive changes $W$, its stationary state, and its relaxation modes, and can
therefore create, suppress, or shift a crossing. Also, for a non-Gibbsian
$\pi_{\rm ness}$, $D(p\Vert\pi_{\rm ness})$ is a Lyapunov measure;
its identification with the physical excess Helmholtz free energy of the
original energies requires additional thermodynamic structure. Our driven
comparison holds the entire generator fixed.

\subsection{A driven three-state example}
\label{sec:nessexample}

To make Eq.~\eqref{eq:identity_ness} concrete we drive the three-state model
of Sec.~\ref{sec:model} around its cycle.  Each forward rate on the cycle
$0\to1\to2\to0$ is multiplied by $e^{F/6}$ and each backward rate by
$e^{-F/6}$, which preserves each edge's geometric-mean kinetic factor while changing
its forward-to-backward ratio. This imposes the cycle affinity
\begin{equation}
\mathcal A=\sum_{\rm cycle}\ln\frac{k_{ij}}{k_{ji}}=F.
\label{eq:affinity}
\end{equation}
We take $F=2$.  The stationary state and the two nonzero relaxation rates of
the driven generator are
\begin{equation}
\begin{gathered}
\pi_{\rm ness}=(0.439083,\,0.548953,\,0.011963),\\
\lambda_2^{\rm ness}=2.04291,\qquad\lambda_3^{\rm ness}=29.84311,
\end{gathered}
\label{eq:ness_numbers}
\end{equation}
both rates are real for this driving, and the housekeeping rate at stationarity is
$\dot e_p^{\rm hk}=\dot e_p^{\rm tot}(\pi_{\rm ness})=0.04256$.

\begin{figure*}[!t]
\centering
\includegraphics[width=\textwidth]{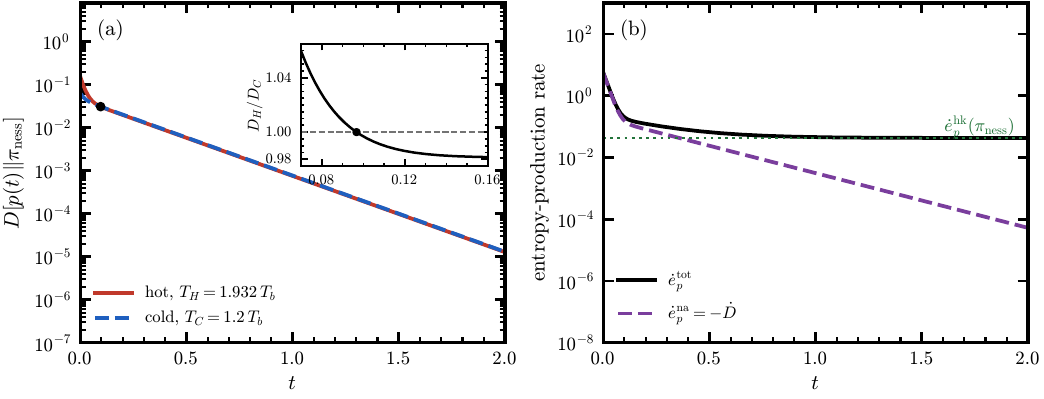}
\caption{Mpemba effect toward a nonequilibrium steady state (driven
three-state cycle, affinity $F=2$).  (a) Relative entropy to
$\pi_{\rm ness}$ for the two preparations of Eq.~\eqref{eq:ness_temps}:
although $D_H(0)>D_C(0)$, the curves cross at $t_M=0.097$; the inset
resolves the ratio $D_H/D_C$ near the crossing.  (b) Along the hot
trajectory the nonadiabatic rate $\dot e_p^{\rm na}=-\dot D$ decays to zero
while the total rate tends to the stationary housekeeping value (dotted
line). The instantaneous gap is $\dot e_p^{\rm hk}(t)$.}
\label{fig:ness}
\end{figure*}

For the preparation we adopt the natural two-step protocol: the system is
equilibrated with the driving \emph{off} at temperature $T_0$, and at $t=0$ the
bath is quenched to $T_b$ and the driving is switched on.  The initial states
are then the canonical distributions \eqref{eq:preps} and the relaxation
proceeds under the driven generator toward $\pi_{\rm ness}$.  We use the fixed initial-temperature pair
\begin{equation}
T_H=1.932\,T_b,\qquad T_C=1.200\,T_b,
\label{eq:ness_temps}
\end{equation}
for which
\begin{equation}
D_H(0)=0.1591734>D_C(0)=0.0646191,
\label{eq:ness_D0}
\end{equation}
a ratio of $2.46$, yet the trajectories cross at
\begin{equation}
t_M=0.0967171 .
\label{eq:ness_tM}
\end{equation}
Figure~\ref{fig:ness}(a) shows the crossing of
$D[p(t)\Vert\pi_{\rm ness}]$; panel (b) compares, along the hot trajectory,
the total entropy production with its nonadiabatic part. The identity
\eqref{eq:identity_ness} is verified independently; at late times
$\dot e_p^{\rm na}\to0$ while $\dot e_p^{\rm tot}$ approaches the stationary
housekeeping rate. Away from stationarity, $\dot e_p^{\rm hk}(t)$ need not
equal that limiting constant. Multistart minimization gives the estimate
$\widehat\rho_{\rm na}\simeq2.03823$; its global optimality is not certified.
The comparison with the ordinary slow eigenvalue $\lambda_2^{\rm ness}=2.04291$ is descriptive,
not a reversible spectral-gap theorem.

\section{Experimental interpretation}
\label{sec:experiment}

The common-bath protocol can be implemented whenever the mesoscopic state of
a system is controllable and its transitions can be sampled. The colloidal
realization of Kumar and Bechhoefer \cite{kumar2020} follows exactly this
logic: the system is prepared in a controlled initial equilibrium
distribution and then relaxes in a fixed final landscape and bath.
Few-state reductions of this kind are standard for mesoscopic systems,
including coupled and polymeric Brownian particles on ratchets and in
thermal landscapes
\cite{Taye2015RectifiedMotionShort,Taye2023TransportingShortPolymer,Taye2025DirectedTransportShort,Taye2026NoiseActivatedDopant,Aragie2014ImpurityDiffusionHarmonic,Aragie2022NoiseFormedTriple,Abebe2022ThermallyActivatedDiffusion,Getahun2008CompetingJumpCycles},
adhesion and unbinding of membranes and strings
\cite{Asfaw2006MembraneAdhesionVia,Asfaw2005AdhesionMultiComponent,Asfaw2007AdhesionInducedLateral,Asfaw2008AdhesionInducedLateral,Asfaw2008UnbindingTransitionsMembranes,Asfaw2007UnbindingTransitionsMulticomponent,Asfaw2011LateralPhaseSeparation,Asfaw2011ExploringElasticFeatures},
and charge transport and fluctuations in condensed matter
\cite{Demeyu2007MonteCarloSimulations,Mahmud2022ThermalQuantumFluctuation}.

For a finite-state realization, the proposed thermodynamic analysis requires
three types of data.  First, equilibrium sampling at the final bath
temperature determines $\pi_{b,i}$.  Second, repeated quenches from two
initial temperatures determine the time-dependent occupation probabilities
$p^{H,C}_i(t)$.  Third, transition counts estimate the rates $k_{ij}$ or,
equivalently, the probability currents.  The relative entropy is then computed
from Eq.~\eqref{eq:Ddef}, while entropy production follows from
Eq.~\eqref{eq:sigma}.  A direct experimental check of the central identity is
\begin{equation}
-\frac{\mathrm d}{\mathrm dt}\,\Dr\;\stackrel{?}{=}\;\dot e_p(t).
\label{eq:expcheck}
\end{equation}
The different relaxation times can then be extracted without fitting a single
exponential.  $\tau_{\rm th}$ is the first time at which $D$ reaches a
prescribed tolerance; $\bar\tau$ is the normalized area under $D(t)$; and
$\tau_{\rm inst}=D/\dot e_p$ can be constructed pointwise.  The spectral time
can be obtained independently from the rate matrix.  The central experimental
signature is therefore not merely that one measured temperature curve falls
faster than another, but that the two copies possess the same post-quench
generator while their preparation-dependent thermodynamic clocks advance
differently.

The bounds add quantitative tests once a common rate matrix has been
estimated. A numerical minimization gives an estimate of $\rho$; a
certified lower bound $\underline\rho$ is needed to assert
$D(t)\le D(0)e^{-2\underline\rho t}$ and
$r(t)\ge2\underline\rho$. For a model with measured mode amplitudes,
Eq.~\eqref{eq:envelopes} supplies an alternative constraint that does not
require the optimal constant. Uncertainty in the measured generator and
preparations must be propagated into any reported bound.  For driven systems the same
program applies with $\pi_{\rm ness}$ estimated from long-time sampling and
$\dot e_p^{\rm na}$ constructed from Eq.~\eqref{eq:signa}; the housekeeping part
then follows by subtraction from the Schnakenberg total.

Finite sampling makes $D$ and $\dot e_p$ noisy near equilibrium.  A practical
analysis should therefore use Bayesian or bootstrap estimates of $p_i(t)$,
regularize very small probabilities, and report confidence intervals for
crossing and threshold times.  Because $\dot e_p$ contains logarithms of forward
and reverse probability fluxes, transitions with few observed events require
particular care.  The endpoint expression
$\Theta(t)=\ln[D(0)/D(t)]$ avoids differentiating $D$ when only the progress
variable is required. Integrating a noisy estimate of $\dot e_p/D$ is not
automatically more stable, especially when $D$ is very small.

\section{Relation to hot water and to the original freezing question}
\label{sec:water}

The present model does not claim that a finite-state Markov chain
accounts for the microscopic reasons why hot water can sometimes freeze
before colder water.  The original water problem can involve evaporation, convection,
supercooling, nucleation, gas content, container contact, and phase-change
kinetics \cite{jeng2006}.  The statistical-mechanical theory instead isolates
a logically prior phenomenon: anomalous relaxation toward a common equilibrium
state can occur even when both systems experience exactly the same post-quench
dynamics.

This separation is useful because it identifies what does not need to change.
In the Markovian Mpemba effect, the hotter preparation does not need a colder
bath, a larger heat-transfer coefficient, or different post-quench rate
constants.  It can relax faster solely because its initial equilibrium
distribution has a more favorable decomposition over the common relaxation
modes.  Water-specific mechanisms may alter the effective generator as the
temperature and phase change; the Markovian theory describes the simpler limit
in which the post-quench law is fixed.

For a macroscopic fluid, the state space is of course enormous.  The
three-state model should therefore be viewed as a minimal proof of principle,
not a molecular model of water.  Its value is to expose the statistical
structure that a more detailed model with fixed post-quench dynamics can reproduce: the initially
hotter ensemble may be farther from equilibrium globally while having
anomalously weak overlap with the slowest route back to equilibrium.

\section{Discussion}
\label{sec:discussion}

The generator determines the available relaxation rates; the preparation
determines their amplitudes. This distinction explains why two systems can
share the same spectral time but reach a fixed thermodynamic tolerance at
different times. Relative entropy adds a thermodynamic interpretation at
equilibrium, and its logarithmic derivative reports the fraction of remaining
excess free energy dissipated per unit time. The clock is a useful reading
of the trajectory, not extra dynamical information.

The global rate bound and the amplitude enclosure use different amounts of
information. The former retains an initial entropy ratio and rate extrema,
so it is inexpensive but discards the small amplitude of the cold state's
slow mode. The latter retains that amplitude and controls the entropy at
finite deviations through an exact scalar inequality. Its much tighter
crossing interval is therefore a quantitative consequence of additional
preparation information. The method is stated explicitly for an exactly
strong, reversible three-state preparation; more modes, degeneracies, and
imperfect tuning require modified envelopes rather than automatic use of
Eqs.~\eqref{eq:loweramp}--\eqref{eq:upperamp}.

Two different limitations should be kept separate. The best modified
logarithmic Sobolev constant has not been globally certified here. All claims
that require a lower certificate use $\kappa/2$, and the main amplitude
enclosure does not use that optimization at all. Separately, a KL crossing
is one measure of anomalous relaxation and need not imply a crossing of mean
energy, temperature, or every monotone distance. Thermomajorization addresses
the stronger ordering question \cite{vu2025}, while resource theories
provide further perspectives on the choice of monotone \cite{summer2026}.

The contribution is consequently focused: an explicit finite-amplitude
crossing calculation and its comparison with globally valid entropy-decay
constraints. Entropy production, functional inequalities, and relaxation
duration are already closely connected in prior work
\cite{bobkov2006,bao2025,auconi2026}, and nonadiabatic dissipation already
governs stationary-state relative-entropy decay \cite{esposito2010}.
The elementary rate bound follows from established entropy-decay estimates.
The static KL--$\chi^2$ comparison used in Sec.~\ref{sec:envelopes} is
also established; the dynamical construction combines it with Markov
contraction and explicit preparation information to bound the first crossing.
Establishing sharp constants across useful families of generators, and
robust crossing intervals under uncertain preparation, are concrete open
problems.

The distinction between a dynamical law and its initial state also motivates
questions beyond thermal relaxation. Stochastic descriptions of
viral dynamics, latent-reservoir reactivation, and viral rebound
\cite{Perelson1993,Perelson1996,Ho1995,Wei1995,NowakBangham1996,Perelson2002,Wodarz2002,Chun1997,Siliciano2003,Murray2016,Li2020,Hill2014,Hill2018,pinkevych2015latency,Pinkevych2019,Wu2020,conway2019rebound,VanDorp2020,fennessey2017barcoded,li2016reservoir,pasternak2020carna,sneller2020kinetics,gunst2025ati,Taye2025CM,Taye2026StochasticFirstPassage,Taye2025DrugWashoutViral,Taye2023HostViralLoad,Taye2020CorrelationBetweenAntiviral,Zhang2021EfficacyAntiviralDrug,Taye2023TimeDependentThermodynamic},
of sedimentation and cellular kinetics
\cite{Taye2019PhysicsErythrocyteSedimentation,Taye2020SedimentationRateErythrocyte,Taye2023DynamicsBloodCells,Mahmud2023ComputationalInvestigationCis},
and thermodynamic descriptions of aging and biological time
\cite{Taye2026NonequilibriumInternalTime,Taye2026BiologicalTimeEvolutionary,Taye2026RelativisticPbteBiological,Taye2026BiologicalProperTime,Taye2026PrincipleBiologicalTime,Taye2026ThermodynamicParametrisationVertebrate,Taye2026NeuralInvestmentEntropy,Taye2026LifetimeCardiacCycle,Taye2026BiologicalTimeEquivalence}
provide possible settings for future comparisons of initial conditions.
These citations supply broader context; no biological or transport
application of the present crossing theorem is established here.
Such an application would first require a justified fixed Markov generator,
a specified stationary reference and distance, and preparations satisfying
the hypotheses of the relevant bound. Whether those assumptions and a
Mpemba-type crossing hold must be checked separately for each model.

\section{Conclusions}
\label{sec:conclusions}

We have compared two canonical preparations under a common final generator
using four operationally distinct relaxation times. The standard balance
$-\dot D=\dot e_p$ makes the accumulated logarithmic entropy reduction a
dimensionless progress variable and yields an exact crossing identity.
Global entropy-decay estimates constrain that identity only when the
constants inserted have the correct bound directions.

For the explicit three-state model we provide an analytic lower certificate
for $\rho$, a proof of the hot trajectory's rate supremum, and a cubic-moment
condition establishing $\rho<\lambda_2$ independently of the numerical
optimization. More usefully, finite-amplitude entropy envelopes retain the
initial mode weights and give an interval-certified crossing enclosure
$0.1659\le t_M\le0.1674$, containing the computed value $0.1662209$.
The hot preparation reaches $D=10^{-5}$ at $t=0.233$, compared with
$t=0.844$ for the cold one, despite
starting $6.12$ times farther from equilibrium in relative entropy.

For a fixed driven generator, nonadiabatic entropy production replaces total
entropy production in the clock balance. Housekeeping dissipation is absent
from that balance, while driving can still affect the generator and the
crossing. These conclusions concern finite-state ensemble relaxation and a
specified thermodynamic measure; they do not supply a theory of water
freezing.

\section*{Data and code availability}

The Python scripts that construct the model and generate the figures, the
machine-readable results, the independent 60-digit check, and the
outward-rounded interval certificate of Appendix~\ref{app:interval} are
provided as Supplemental Material, together with a README that lists the
commands and software versions needed to reproduce every reported number.
All numbers refer to the exact parameter definitions rather than the rounded
matrix printed in Eq.~\eqref{eq:Wmatrix}. Numerical estimates and analytic
bound evaluations are identified separately.

\appendix

\section{Orthogonality of the relaxation modes}
\label{app:selfadjoint}

The relative deviation $f_i=p_i/\pi_{b,i}-1$ obeys $\dot f=Lf$, where
\begin{equation}
(Lf)_i=\sum_{j\ne i}k_{ij}(f_j-f_i),
\qquad W=\operatorname{diag}(\pi_b)\,L\,\operatorname{diag}(\pi_b)^{-1}.
\label{eq:Lexplicit}
\end{equation}
With the equilibrium inner product
$\langle u,v\rangle_{\pi_b}=\sum_i\pi_{b,i}u_iv_i$, detailed balance,
$\pi_{b,i}k_{ij}=\pi_{b,j}k_{ji}$, gives
\begin{equation}
\langle u,Lv\rangle_{\pi_b}
=-\frac12\sum_{i,j}\pi_{b,i}k_{ij}(u_i-u_j)(v_i-v_j)
=\langle Lu,v\rangle_{\pi_b},
\label{eq:selfadjoint}
\end{equation}
so $L$ is self-adjoint in this inner product. Its eigenvalues
$-\lambda_n$ are real and nonpositive, and its eigenfunctions $\phi_n$ can be
chosen orthonormal, $\langle\phi_m,\phi_n\rangle_{\pi_b}=\delta_{mn}$, with
$\phi_1=1$. The vectors $v_n=\operatorname{diag}(\pi_b)\phi_n$ then satisfy
$Wv_n=-\lambda_nv_n$ and
$\sum_iv_{m,i}v_{n,i}/\pi_{b,i}=\langle\phi_m,\phi_n\rangle_{\pi_b}
=\delta_{mn}$, which is Eq.~\eqref{eq:orth}. Taking the inner product of
$f(0)$ with $\phi_n$ gives the amplitudes $b_n$. The quadratic form in
Eq.~\eqref{eq:selfadjoint}, $\mathcal E(f,f)=-\langle f,Lf\rangle_{\pi_b}$,
is the Dirichlet form used in Sec.~\ref{sec:bounds} and
Appendix~\ref{app:unique_crossing}.

\section{Entropy-dissipation identity under detailed balance}
\label{app:identity}

Starting from
\begin{equation}
D=\sum_ip_i\ln\frac{p_i}{\pi_i},
\label{eq:appA_D}
\end{equation}
normalization gives
\begin{equation}
\dot D=\sum_i\dot p_i\ln\frac{p_i}{\pi_i}.
\label{eq:appA_Ddot}
\end{equation}
Using the master equation in current form,
\begin{equation}
\dot p_i=-\sum_jJ_{ij},
\label{eq:appA_master}
\end{equation}
with $J_{ij}=-J_{ji}$,
\begin{equation}
\dot D=-\frac12\sum_{i,j}J_{ij}
\left[\ln\frac{p_i}{\pi_i}-\ln\frac{p_j}{\pi_j}\right].
\label{eq:appA_pair}
\end{equation}
Detailed balance implies $\pi_ik_{ij}=\pi_jk_{ji}$, so
\begin{equation}
\ln\frac{p_i/\pi_i}{p_j/\pi_j}=\ln\frac{k_{ij}p_i}{k_{ji}p_j}.
\label{eq:appA_db}
\end{equation}
Therefore
\begin{equation}
\dot D=-\frac12\sum_{i,j}J_{ij}\ln\frac{k_{ij}p_i}{k_{ji}p_j}=-\dot e_p.
\label{eq:appA_final}
\end{equation}
Since each pair contribution $(a-b)\ln(a/b)$ is nonnegative, $\dot e_p\ge0$ and
$D$ decreases monotonically.

\section{Quadratic expansion of the relative entropy}
\label{app:quadratic}

Write
\begin{equation}
p_i=\pi_i(1+f_i),\qquad \sum_i\pi_if_i=0.
\label{eq:appB_f}
\end{equation}
Then
\begin{align}
D&=\sum_i\pi_i(1+f_i)\ln(1+f_i)
\label{eq:appB_D}\\
&=\sum_i\pi_i\left[f_i+\frac12f_i^2-\frac16f_i^3+O(f_i^4)\right].
\label{eq:appB_expand}
\end{align}
The linear term vanishes by normalization, hence
\begin{equation}
D=\frac12\langle f,f\rangle_\pi+O(f^3).
\label{eq:appB_final}
\end{equation}
Substituting the reversible mode expansion in Eq.~\eqref{eq:fexp} gives
Eq.~\eqref{eq:Dmodes}.  Differentiation gives Eq.~\eqref{eq:sigmamodes}.

\section{Integrated relaxation time in the quadratic regime}
\label{app:integrated}

Define $D_{\rm quad}(t)=\tfrac12\sum_{n\ge2}b_n^2e^{-2\lambda_nt}$. Then
\begin{equation}
\int_0^\infty D_{\rm quad}(t)\,\mathrm dt
=\frac12\sum_{n\ge2}b_n^2\int_0^\infty e^{-2\lambda_nt}\,\mathrm dt
=\frac14\sum_{n\ge2}\frac{b_n^2}{\lambda_n}.
\label{eq:appC_int}
\end{equation}
Meanwhile
\begin{equation}
D_{\rm quad}(0)=\frac12\sum_{n\ge2}b_n^2.
\label{eq:appC_D0}
\end{equation}
Their ratio is
\begin{equation}
\bar\tau_{\rm quad}=\sum_{n\ge2}\frac{w_n}{2\lambda_n},
\qquad
w_n=\frac{b_n^2}{\sum_{m\ge2}b_m^2}.
\label{eq:appC_tau}
\end{equation}
This also yields the bounds
\begin{equation}
\frac{1}{2\lambda_N}\le\bar\tau_{\rm quad}\le\frac{1}{2\lambda_2}
\label{eq:appC_bounds}
\end{equation}
for the quadratic approximation. For a fixed generator and an initial
perturbation uniformly small in $\|f(0)\|_\infty$, the exact quantity
satisfies $\bar\tau=\bar\tau_{\rm quad}+O(\|f(0)\|_\infty)$; it need
not obey these exact endpoint inequalities.

\section{Numerical construction and independent checks}
\label{app:numerics}

\subsection{Reproducible model and stable entropy evaluation}

The rates are constructed directly from Eqs.~\eqref{eq:energies},
\eqref{eq:rates}, and \eqref{eq:gammas}, rather than from the rounded matrix
in the text. For detailed balance the symmetric matrix
\begin{equation}
S=\operatorname{diag}(\pi_b)^{-1/2}W
\operatorname{diag}(\pi_b)^{1/2}
\label{eq:appD_sym}
\end{equation}
is diagonalized with an orthonormal eigensolver. If $w_n$ is a normalized
eigenvector of $S$, then $v_{n,i}=\sqrt{\pi_{b,i}}\,w_{n,i}$,
$\phi_{n,i}=w_{n,i}/\sqrt{\pi_{b,i}}$, and
$b_n=\sum_i[p_i(0)-\pi_{b,i}]\phi_{n,i}$. Propagation uses
\begin{equation}
f_i(t)=\sum_{n\ge2}b_n e^{-\lambda_nt}\phi_{n,i},
\qquad p_i(t)=\pi_{b,i}[1+f_i(t)].
\label{eq:appD_evolve}
\end{equation}
No numerical time step is needed. In the driven case a stationary linear
solve and the two nonstationary right eigenvectors give the corresponding
deviation evolution; a separate matrix exponential checks it.

Directly summing $p_i\ln(p_i/\pi_i)$ can suffer cancellation when $p$ is
very close to $\pi$. We instead use normalization to evaluate
\begin{equation}
D=\sum_i\pi_i g(f_i),\quad
g(x)=\sum_{n=2}^{\infty}\frac{(-1)^n x^n}{n(n-1)},
\label{eq:appD_stable}
\end{equation}
The power-series representation holds for $|x|<1$.
For $|x|<10^{-3}$, terms through $n=12$ are used; otherwise the logarithm
is evaluated by \texttt{log1p}. The discarded series remainder is bounded
by $|x|^{13}/[156(1-|x|)]$. The detailed-balance flux is evaluated as
$\pi_i k_{ij}(f_i-f_j)$, and the associated logarithm as
$\ln(1+f_i)-\ln(1+f_j)$. These expressions avoid subtracting two nearly
equal equilibrium fluxes. The modal derivative separately gives
\begin{equation}
-\dot D=-\sum_i\pi_i\dot f_i\ln(1+f_i),
\label{eq:appD_derivative}
\end{equation}
which is checked against the flux expression.

\subsection{Root finding, optimization, and ideal preparation}

The strong temperature is the nontrivial root of $b_2(T)=0$ bracketed by
$2<T<4$. The double-precision root has a residual slow coefficient below
$10^{-16}$. For the ideal strong trajectory, that coefficient is set to zero
by the mathematical definition of the preparation; this changes the initial
probabilities by less than $10^{-15}$. A rounded-temperature experiment
would retain a small residual, with the late crossover described in
Eq.~\eqref{eq:tuning}.

The relative-entropy crossing and threshold times are roots of
$D_H(t)-D_C(t)$ and $D_\alpha(t)-\varepsilon$, respectively.
Adaptive quadrature evaluates the integrated times. The search for
$\widehat\rho$ uses 4000 Dirichlet samples with fixed seed 3, an interior
mesh, and 12 local Nelder--Mead starts; the driven search uses seed 7.
This reproducible procedure is a search for low values of the functional,
not a proof of its global minimum. The certified comparison constant
$\kappa$ is computed separately from Eq.~\eqref{eq:resetbound}.
The NESS temperatures are fixed at exactly $1.932$ and $1.2$ in both the
results and figures.

\subsection{Independent validation and precision}

The modal propagation is compared with a matrix exponential at six times
for both preparations and both generators. The maximum probability
difference is below $6\times10^{-16}$, and normalization errors are below
$3\times10^{-16}$. The flux and derivative evaluations of entropy
dissipation agree to an absolute error below $3\times10^{-15}$; the NESS
production decomposition agrees below $4\times10^{-15}$. An independent
centered finite difference away from very small entropies agrees with
$-\dot D$ within $8\times10^{-10}$. These are numerical checks over the
reported set of states, not substitutes for the analytic identities.

A separate implementation uses 60-digit decimal arithmetic, the analytic
characteristic quadratic for the nonzero eigenvalues, cross-product
eigenvectors, and bisection. It uses neither the double-precision eigensolver
nor its root finder. It reproduces
\begin{align}
t_M&=0.16622089014064838748\ldots,\\
\ell(s_4)&=0.16594971729196675519\ldots,\\
u(s_4)&=0.16738643933850475739\ldots.
\end{align}
The ordinary and high-precision results agree well beyond the digits quoted
in the text. High-precision point evaluation alone does not constitute a
directed interval-arithmetic proof. The separate outward-rounded
calculation in Appendix~\ref{app:interval} certifies the displayed crossing
enclosure and the numerical signs needed by the analytic arguments.
None of these computations certifies global optimality of $\widehat\rho$.
The source package identifies these distinct levels of evidence explicitly.

\section{Proofs of the global entropy and rate bounds}
\label{app:bounds}

For $p(t)\ne\pi$, the definition of $\rho$ implies
$\dot e_p(p(t))\ge2\rho D(p(t)\Vert\pi)$. Combining this with
$-\dot D=\dot e_p$ gives
\begin{equation}
\frac{\mathrm d}{\mathrm dt}\{e^{2\rho t}D(t)\}\le0.
\end{equation}
Integrating gives $D(t)\le D(0)e^{-2\rho t}$. Integrating that bound over
$t\ge0$ and dividing by $D(0)$ gives the integrated-time bound. If
$0<\varepsilon<D(0)$, the exponential envelope reaches $\varepsilon$ at
$\ln[D(0)/\varepsilon]/(2\rho)$, by which time the exact trajectory must
also have reached the threshold. If the initial state already meets the
threshold, its time is zero. This proves Proposition 1 and its corollaries.
Proposition 2 follows from Eq.~\eqref{eq:rateproof} by integrating the
pointwise bounds on the two rates.

For completeness, in a finite irreducible reversible chain the entropy
production is positive at every nonstationary interior state. At a
nonstationary boundary state at least one positive-probability state sends
current to a zero-probability state, causing entropy production to diverge.
Near equilibrium the ratio $\dot e_p/(2D)$ has a positive quadratic lower
limit set by $\lambda_2$. Compactness away from these two limits therefore
ensures $\rho>0$. For an interior nonstationary trajectory the rate is
continuous on bounded time intervals and approaches twice its slowest
populated eigenvalue at long times; hence its trajectory supremum is finite.

\subsection{A certified lower bound on the entropy-decay constant}
\label{app:reset}

The variational number $\rho=\inf_{p\ne\pi}\dot e_p(p)/(2D(p\Vert\pi))$
is a property of the generator. A state found by numerical minimization
provides an \emph{upper} bound on this infimum, not the lower bound needed
to certify an entropy-decay estimate. We therefore separate the numerical
estimate from a conservative analytic bound.

For a complete transition graph, define
\begin{equation}
\kappa=\min_{i\ne j}\frac{k_{ij}}{\pi_j}>0,
\qquad \widetilde k_{ij}=k_{ij}-\kappa\pi_j\ge0.
\label{audit:kappa}
\end{equation}
The original process is the sum of a reset process, with rates
$k^{\rm reset}_{ij}=\kappa\pi_j$, and a residual Markov process with rates
$\widetilde k_{ij}$. Both have stationary distribution $\pi$. If the
original process is reversible, the residual process is also reversible:
\begin{equation}
\pi_i\widetilde k_{ij}
=\pi_i k_{ij}-\kappa\pi_i\pi_j
=\pi_j\widetilde k_{ji}.
\end{equation}
Under the reset dynamics, $\dot p_i=\kappa(\pi_i-p_i)$, and hence
\begin{align}
\dot e_p^{\rm reset}
&=-\sum_i\dot p_i\ln\frac{p_i}{\pi_i}
=\kappa\sum_i(p_i-\pi_i)\ln\frac{p_i}{\pi_i}\\
&=\kappa\{D(p\Vert\pi)+D(\pi\Vert p)\}
\ge\kappa D(p\Vert\pi).
\end{align}
Entropy dissipation is linear in the generator. The residual process has
nonnegative entropy dissipation, so
\begin{equation}
\dot e_p(p)\ge\kappa D(p\Vert\pi),\qquad
\boxed{\rho\ge\underline\rho:=\kappa/2.}
\label{audit:rholower}
\end{equation}
For the equilibrium example, $\kappa=1.1298931023$ and therefore
$\underline\rho=0.5649465511$. This bound is deliberately conservative;
it certifies $D(t)\le D(0)e^{-\kappa t}$ without assuming that a numerical
search has found the global minimizer. The same construction applies to
nonadiabatic entropy dissipation for a fixed nonreversible generator;
reversibility of the residual generator is unnecessary in that case.
The driven example has $\kappa_{\rm na}=0.5747299078$.

\subsection{When the entropy-decay constant is strictly below the spectral gap}
\label{app:cubic}

Let $\phi$ be a normalized real eigenfunction with eigenvalue $-\lambda_2$,
so that $\langle\phi\rangle_\pi=0$ and $\langle\phi^2\rangle_\pi=1$.
For sufficiently small positive or negative $\epsilon$, the state
$p_i=\pi_i(1+\epsilon\phi_i)$ has positive components. Define
$\mu_3=\langle\phi^3\rangle_\pi$. Direct expansion gives
\begin{equation}
D(\epsilon)=\frac{\epsilon^2}{2}
-\frac{\mu_3\epsilon^3}{6}+O(\epsilon^4).
\end{equation}
Because this state relaxes along one eigenmode,
$\dot\epsilon=-\lambda_2\epsilon$. Thus
\begin{equation}
\dot e_p(\epsilon)
=\lambda_2\epsilon\frac{dD}{d\epsilon}
=\lambda_2\epsilon^2
-\frac{\lambda_2\mu_3\epsilon^3}{2}+O(\epsilon^4),
\end{equation}
and division yields
\begin{equation}
\boxed{\frac{\dot e_p(\epsilon)}{2D(\epsilon)}
=\lambda_2\left[1-\frac{\mu_3\epsilon}{6}
+O(\epsilon^2)\right].}
\label{audit:cubicratio}
\end{equation}
Consequently, $\mu_3\ne0$ is a sufficient condition for
$\rho<\lambda_2$: choose $\epsilon$ with the same sign as $\mu_3$ and
small enough that the linear correction controls the remainder. In the
example, the normalized eigenfunction is approximately
\begin{equation}
\phi\simeq(-0.79722822,1.25141264,-0.83593505),
\end{equation}
which gives $\mu_3\simeq0.4522295539\ne0$.
Appendix~\ref{app:interval} certifies $0.4522<\mu_3<0.4523$.
Strict inequality is therefore established
analytically for this model, independently of the reported numerical
minimization. If the slow eigenspace is degenerate, it is enough that one
normalized function in that eigenspace has a nonzero third moment.

If $\mu_3=0$, the calculation gives no strictness conclusion. For example,
for a symmetric two-state chain with both jump rates equal to $k$,
$\pi=(1/2,1/2)$ and $\lambda_2=2k$. Writing
$p=((1+x)/2,(1-x)/2)$ gives
\begin{equation}
D(x)=\sum_{n=1}^{\infty}\frac{x^{2n}}{2n(2n-1)},
\qquad
\dot e_p(x)=\lambda_2 xD'(x).
\end{equation}
Every coefficient is nonnegative, and $xD'(x)\ge2D(x)$. Hence
$\dot e_p/(2D)\ge\lambda_2$ for all $|x|<1$, with equality approached as
$x\to0$. This chain has $\rho=\lambda_2$.

\subsection{Exact peak rate for the strong-Mpemba preparation}
\label{app:hotcap}

For the selected three-state preparation, the slow mode is absent. Its
relative deviation is exactly
\begin{equation}
\begin{gathered}
\frac{p_i^H(t)}{\pi_i}-1=x(t)\psi_i,\qquad x(t)=e^{-\lambda_3t},\\
\psi\simeq(-0.22289599,0.00616613,4.39396646).
\end{gathered}
\end{equation}
The entries shown are rounded; the calculation uses the exact rates and the
root defining $T_\star$. Define
\begin{equation}
H(x)=\sum_i\pi_i\{(1+x\psi_i)\ln(1+x\psi_i)-x\psi_i\}.
\end{equation}
Then $D_H(t)=H(x(t))$ and
\begin{equation}
r_H(t)=\lambda_3\frac{xH'(x)}{H(x)}.
\end{equation}
To bound this expression for the entire trajectory, calculate
\begin{align}
H'''(x)&=-S(x),\qquad
S(x)=\sum_i\frac{\pi_i \psi_i^3}{(1+x\psi_i)^2},\\
S'(x)&=-2\sum_i\frac{\pi_i \psi_i^4}{(1+x\psi_i)^3}<0.
\end{align}
In this example $S(1)\simeq0.0737338378>0$;
Appendix~\ref{app:interval} certifies $0.0737<S(1)<0.0738$.
Since $S$ decreases with $x$,
$S(x)\ge S(1)>0$ throughout $0\le x\le1$, so $H'''(x)<0$.
Now set $G(x)=2H(x)-xH'(x)$. Its derivatives obey
\begin{equation}
G'(x)=H'(x)-xH''(x),\qquad
G''(x)=-xH'''(x)\ge0.
\end{equation}
Because $G(0)=G'(0)=0$, first $G'(x)\ge0$ and then $G(x)\ge0$.
Therefore $xH'(x)\le2H(x)$ and
\begin{equation}
r_H(t)\le2\lambda_3.
\end{equation}
The small-$x$ expansion $H(x)=\frac12\langle \psi^2\rangle_\pi x^2+O(x^3)$
shows that equality is approached as $t\to\infty$. The exact supremum is
\begin{equation}
\boxed{R_H=2\lambda_3\simeq44.2025176815.}
\label{audit:RH}
\end{equation}
It is a limiting value, not a maximum attained at the endpoint of a
finite numerical sampling window. Combining this result with
Eq.~\eqref{audit:rholower} certifies the conservative crossing estimate
\begin{equation}
t_M\ge\frac{\ln[D_H(0)/D_C(0)]}{2\lambda_3-\kappa}
=0.04206482\ldots.
\label{audit:coarsecross}
\end{equation}
Substitution of an estimated global minimum would give approximately
$0.04515$, but that number is conditional on the minimization estimate
and is not the certified bound reported here.

\section{A spectral enclosure for the first entropy crossing}
\label{app:envelopes}

The global rate bound discards the cold preparation's small slow-mode
amplitude. Retaining the amplitudes leads to a much sharper result without
requiring a global optimization. Write
\begin{equation}
\begin{gathered}
f_i^\alpha(t)=\frac{p_i^\alpha(t)}{\pi_i}-1,\qquad
Q_\alpha(t)=\sum_i\pi_i[f_i^\alpha(t)]^2,\\
m_\alpha(s)=\min_i f_i^\alpha(s),\qquad
M_\alpha(s)=\max_i f_i^\alpha(s).
\end{gathered}
\end{equation}
The scalar function
\begin{equation}
F(x)=\frac{(1+x)\ln(1+x)-x}{x^2}
=\int_0^1\frac{1-y}{1+yx}\,\mathrm dy,
\label{audit:F}
\end{equation}
with $F(0)=1/2$, is positive and decreasing on $x>-1$; differentiation under the integral
gives $F'(x)=-\int_0^1y(1-y)/(1+yx)^2\,\mathrm dy<0$.
As shown in Sec.~\ref{sec:envelopes}, $f_i(t)=\sum_jK_{ij}f_j(s)$ with
nonnegative weights $K_{ij}$ that sum to one over $j$. Every later value is therefore a convex combination of values at
time $s$, giving
\begin{equation}
m_\alpha(s)\le f_i^\alpha(t)\le M_\alpha(s),\qquad t\ge s.
\end{equation}
Normalization cancels the linear term in the entropy, so
\begin{align}
D_\alpha(t)
&=\sum_i\pi_i[f_i^\alpha(t)]^2F(f_i^\alpha(t)),\\
F(M_\alpha(s))Q_\alpha(t)
&\le D_\alpha(t)\le F(m_\alpha(s))Q_\alpha(t),\qquad t\ge s.
\label{audit:entropy-envelopes}
\end{align}
These inequalities are exact at finite amplitude; they are not a quadratic
truncation.

For the three-state strong-Mpemba pair, orthonormality gives
\begin{equation}
Q_H(t)=Ae^{-2\lambda_3t},\qquad
Q_C(t)=Be^{-2\lambda_2t}+Ce^{-2\lambda_3t},
\end{equation}
where
\begin{equation}
\begin{aligned}
A&=b_{3,H}^2\simeq0.5877216042,\\
B&=b_{2,C}^2\simeq0.0006409074,\\
C&=b_{3,C}^2\simeq0.0667175704.
\end{aligned}
\end{equation}
Suppose a previous argument has established that the first crossing
$t_M\ge s$. At that crossing, the lower envelope for $D_H$ cannot exceed
the upper envelope for $D_C$. Multiplying by $e^{2\lambda_3t_M}$ yields
\begin{equation}
F(M_H(s))A\le F(m_C(s))
\{Be^{2(\lambda_3-\lambda_2)t_M}+C\}.
\end{equation}
Whenever the logarithm's argument is positive, this implies
\begin{equation}
t_M\ge \ell(s)=
\frac{1}{2\Delta}
\ln\!\left[\frac{A F(M_H(s))/F(m_C(s))-C}{B}\right],
\label{audit:L}
\end{equation}
For an upper bound, use the upper envelope for $D_H$ and the lower envelope
for $D_C$. At any $t\ge s$ for which
\begin{equation}
F(m_H(s))A\le F(M_C(s))
\{Be^{2(\lambda_3-\lambda_2)t}+C\},
\end{equation}
one has $D_H(t)\le D_C(t)$. Since $D_H(0)>D_C(0)$ and the trajectories
are continuous, the first crossing has already occurred by that time.
Thus, if the expression is defined and $u(s)\ge s$,
\begin{equation}
t_M\le u(s)=
\frac{1}{2\Delta}
\ln\!\left[\frac{A F(m_H(s))/F(M_C(s))-C}{B}\right],
\label{audit:U}
\end{equation}

Starting from $s_0=0$ and applying the lower bound repeatedly gives
Table~\ref{tab:envelope_iterations}. At each step the previously proved
lower bound is retained if the new expression is smaller; the best upper
bound is retained as well. The five-step values in the independent
calculation of Appendix~\ref{app:numerics} yield
Eq.~\eqref{eq:tMrefined_num}. No optimization of $\rho$ enters this
argument.

\subsection{Uniqueness of the crossing in the numerical example}
\label{app:unique_crossing}

The envelope construction bounds the first meeting without presupposing
uniqueness. For the selected pair, a separate rate comparison proves that
the ordering changes only once. Let
\begin{equation}
\mathcal E(f,f)=\frac12\sum_{i,j}\pi_{b,i}k_{ij}(f_i-f_j)^2,
\quad
\Lambda_\alpha(t)=\frac{\mathcal E(f^\alpha,f^\alpha)}{Q_\alpha(t)}.
\label{eq:unique_rayleigh}
\end{equation}
For two values $a,b$ in the interval $[m_\alpha(s),M_\alpha(s)]$,
the mean-value theorem gives
\begin{equation}
\frac{(a-b)^2}{1+M_\alpha(s)}
\le(a-b)\{\ln(1+a)-\ln(1+b)\}
\le\frac{(a-b)^2}{1+m_\alpha(s)}.
\end{equation}
Applying this inequality to every edge in the dissipation form, and then
using Eq.~\eqref{eq:envelopes} to bound its entropy denominator, yields
\begin{equation}
\frac{\Lambda_\alpha(t)}
{[1+M_\alpha(s)]F(m_\alpha(s))}
\le r_\alpha(t)\le
\frac{\Lambda_\alpha(t)}
{[1+m_\alpha(s)]F(M_\alpha(s))}
\label{eq:unique_rate_envelopes}
\end{equation}
for all $t\ge s$. The pure fast mode has $\Lambda_H(t)=\lambda_3$. For the cold preparation,
\begin{equation}
\Lambda_C(t)=\frac{\lambda_2+\lambda_3 q(t)}{1+q(t)},
\qquad q(t)=\frac CB e^{-2(\lambda_3-\lambda_2)t},
\end{equation}
so
\begin{equation}
\frac{d\Lambda_C}{dt}
=-\frac{2(\lambda_3-\lambda_2)^2q(t)}{[1+q(t)]^2}\le0.
\end{equation}
Consequently the hot lower bound in Eq.~\eqref{eq:unique_rate_envelopes}
is constant, and its cold upper bound is largest at the reference time.
Take $s=0.1444$, which is below the already established lower bound
$t_M\ge0.14445557\ldots$. Evaluation gives
\begin{equation}
\begin{aligned}
\frac{\lambda_3}{[1+M_H(s)]F(m_H(s))}&=37.32457\ldots,\\
\frac{\Lambda_C(s)}{[1+m_C(s)]F(M_C(s))}&=14.25613\ldots.
\end{aligned}
\end{equation}
Thus conservatively $r_H(t)>37$ and $r_C(t)<15$ for all $t\ge s$.
The numerical separation is certified by interval evaluation in
Appendix~\ref{app:interval}.
Equation~\eqref{eq:ratio} then implies that $D_H(t)/D_C(t)$ is strictly
decreasing on this whole interval. Its value exceeds one before the
first crossing and tends to zero because the hot preparation has no slow
mode while the cold preparation does. There is exactly one crossing,
and $D_H(t)<D_C(t)$ at every later finite time.

\section{Positivity and scope of nonadiabatic and housekeeping production}
\label{app:ness}

Let $k_{ij}$ be fixed in time, let $\pi$ be a positive stationary
distribution, and assume the transition graph is irreducible. Put
$h_i=p_i/\pi_i$. Differentiating the relative entropy and using
normalization gives
\begin{equation}
-\dot D=\sum_{i,j}\pi_i k_{ij}h_i\ln\frac{h_i}{h_j}.
\end{equation}
Stationarity implies
$\sum_{i,j}\pi_i k_{ij}(h_i-h_j)=0$. We can therefore rewrite the
nonadiabatic entropy-production rate as
\begin{equation}
\dot e_p^{\rm na}=-\dot D
=\sum_{i,j}\pi_i k_{ij}
\left[h_i\ln\frac{h_i}{h_j}-h_i+h_j\right]\ge0.
\label{audit:na-positive}
\end{equation}
Every term is nonnegative because $a\ln(a/b)-a+b\ge0$ for positive
$a,b$. Equality requires equal $h$ on every allowed edge; irreducibility
then gives $p=\pi$.

All sums in this appendix run over allowed off-diagonal transitions.
For finite total entropy production, assume dynamic reversibility,
$k_{ij}>0$ if and only if $k_{ji}>0$. Define the stationary dual rates
$k^*_{ij}=\pi_jk_{ji}/\pi_i$. Stationarity gives equal total exit rates,
$\sum_jk^*_{ij}=\sum_jk_{ij}$. The housekeeping contribution is
\begin{align}
\dot e_p^{\rm hk}
&=\sum_{i,j}p_i k_{ij}\ln\frac{k_{ij}}{k^*_{ij}}\\
&=\sum_{i,j}p_i
\left[k_{ij}\ln\frac{k_{ij}}{k^*_{ij}}-k_{ij}+k^*_{ij}\right]\ge0.
\end{align}
It follows that $\dot e_p^{\rm tot}=\dot e_p^{\rm na}+\dot e_p^{\rm hk}$.
Only $\dot e_p^{\rm na}$ appears in $-\dot D$ for a fixed generator and
stationary reference. This does not make driving irrelevant: a change in
driving generally changes $k_{ij}$, $\pi$, the relaxation modes, and the
crossing itself. Also, the phrase ``nonadiabatic entropy production'' is
preferable to an unqualified ``excess entropy production'', which can be
confused with excess environmental entropy flow.

For time-dependent $\pi(t)$, the additional reference term is explicit:
\begin{equation}
\frac{d}{dt}D[p(t)\Vert\pi(t)]
=\sum_i\dot p_i\ln\frac{p_i}{\pi_i}
-\sum_i p_i\,\frac{d}{dt}\ln\pi_i.
\end{equation}
The fixed-reference clock identity must not be applied unchanged to such
protocols. Reversible spectral formulas likewise do not extend verbatim to
a nonreversible generator, whose modes need not be orthogonal or even real.

\section{Outward-rounded certification of the three-state example}
\label{app:interval}

The analytic proofs above reduce the example to a finite number of
numerical inequalities.  We now certify those inequalities, including the
displayed crossing interval, by outward-rounded interval arithmetic.  The
certificate concerns the exact energies and kinetic prefactors in
Eqs.~\eqref{eq:energies}, \eqref{eq:rates}, and \eqref{eq:gammas}, with
$T_b=1$, and the ideal strong preparation.  It does not certify the global
minimization that defines $\rho$, or account for uncertainty in measured
model parameters.

\subsection{Arithmetic and the strong-preparation root}

The supplied script \texttt{interval\_certificate.py} uses only Python's
standard-library \texttt{decimal} module, at 90 significant digits.
All input decimals are exact rationals.  Addition, subtraction,
multiplication, and division round lower endpoints toward $-\infty$ and
upper endpoints toward $+\infty$.  For the increasing functions
$\exp$, $\ln$, and $\sqrt{\phantom{x}}$, each endpoint evaluation is
rounded to nearest and then enlarged by one representable number in the
required direction.  The correctly rounded elementary operations and the
predecessor/successor operations have documented arithmetic semantics.\footnote{
Python Software Foundation,
\href{https://docs.python.org/3.12/library/decimal.html}{\texttt{decimal}
documentation (Python 3.12)};
\href{https://speleotrove.com/decimal/daops.html}{General Decimal Arithmetic
Specification}, version 1.70, arithmetic operations.}
Consequently each interval contains the corresponding exact real value;
increasing the working precision alone is not the certification argument.

For completeness, let $L$ be the backward generator in
Eq.~\eqref{eq:Lexplicit}.  Define
\begin{align}
a&=-\operatorname{tr}L,\qquad
c=\sum_{i<j}(L_{ii}L_{jj}-L_{ij}L_{ji}),\\
\lambda_{2,3}&=\frac{a\mp\sqrt{a^2-4c}}{2}.
\label{eq:interval_spectrum}
\end{align}
The zero eigenvalue is exact by probability conservation.  The remaining
two eigenvalues are enclosed by evaluating these formulas with intervals.
A cross product of the first two rows of $L+\lambda_n I$ gives a
nonzero eigenfunction; its squared $\pi_b$ norm has a strictly positive
lower bound, so normalization is valid throughout the enclosure.  The
orientations are chosen so that $\phi_{2,1}>0$ and $\phi_{3,2}>0$, with
state labels $0,1,2$.

The strong root is bracketed by sign-tested bisection of
$b_2(T)=\sum_i\pi_i(T)\phi_{2,i}$ starting from $[2,4]$.
Every accepted endpoint has an interval value of $b_2$ lying strictly on
one side of zero.  Continuity therefore preserves a root in the bracket.
Its derivative is also enclosed using
\begin{equation}
b_2'(T)=\frac{\sum_i E_i\phi_{2,i}\pi_i(T)-U(T)b_2(T)}{T^2}.
\end{equation}
The derivative is strictly negative on the final bracket, so that bracket
contains exactly one root.  An outward-rounded presentation of it is
\begin{equation}
\begin{split}
T_\star\in[&2.82342865666183523698,\\
           &2.82342865666183523699].
\end{split}
\label{eq:interval_Tstar}
\end{equation}

\subsection{Certified conditions and crossing bounds}

Using this root enclosure, interval evaluation gives the following
deliberately widened bounds:
\begin{align}
2.0540625459&<\lambda_2<2.0540625461,\\
22.1012588407&<\lambda_3<22.1012588408,\\
0.4522&<\mu_3<0.4523,\\
0.0737&<S(1)<0.0738,\\
1.1298931022&<\kappa<1.1298931024.
\label{eq:interval_signs}
\end{align}
Thus the hypotheses used in Appendix~\ref{app:cubic} and
Appendix~\ref{app:hotcap} are certified: $\mu_3\ne0$ and $S(1)>0$.
They establish $\rho<\lambda_2$ and $R_H=2\lambda_3$ for this exact
model.  The same calculation gives a valid conservative crossing bound
$t_M>0.04206482$ from the reset certificate.

For the amplitude construction, $A$, $B$, $C$, and the values of $F$ are
all interval evaluated.  The exact monotonicity of $F$ is used to enclose
its value on an interval by its endpoint values.  Starting at $s_0=0$,
each new reference is the lower endpoint of the preceding lower-bound
interval, rounded further downward to 30 decimal places.  This makes
$s_n\le t_M$ an inductive certificate, rather than an assumption about
rounded iterates.  Five steps give
\begin{align}
0.1659497172&<\ell(s_4)<0.1659497174,\\
0.1673864392&<u(s_4)<0.1673864395.
\end{align}
The slight downward rounding of the reference times does not affect the
digits reported in Table~\ref{tab:envelope_iterations}.  In particular,
\begin{equation}
\boxed{0.1659<t_M<0.1674}
\label{eq:interval_crossing}
\end{equation}
is now a certified enclosure, not merely a high-precision evaluation of
the analytic formulas.

Finally, at the exact reference $s=0.1444$, interval evaluation of the
two rate-envelope expressions in Appendix~\ref{app:unique_crossing}
gives a hot lower bound in $[37.3245,37.3247]$ and a cold upper bound
in $[14.2560,14.2563]$.  Their separation certifies $r_H(t)>37$ and
$r_C(t)<15$ for every $t\ge s$.  Since the first crossing is later than
$s$, this completes the uniqueness argument.  Direct interval evaluation
also confirms $D_H-D_C>0$ at $t=0.1659$ and $D_H-D_C<0$ at $t=0.1674$;
these checks are independent of the iterative construction of the bounds.

The accompanying JSON certificate records the full endpoint intervals,
root signs, and each iteration. These
computer-assisted certificates complement the independent 60-digit
point calculation in Appendix~\ref{app:numerics}.  Neither computation
establishes global optimality of $\widehat\rho$.

\end{document}